\documentclass[letterhead,12pt]{article}
\pdfoutput=1
\usepackage{jheppub}

\usepackage{amsfonts} 
\usepackage{amsmath}
\usepackage{amsthm}
\usepackage{graphicx}
\usepackage{youngtab}
\usepackage{latexsym}
\newcommand{\p}{\partial}

\title{A giant graviton genealogy}

\author[a]{Yolanda Lozano,}
\author[b]{~Jeff Murugan,}
\author[c]{~Andrea Prinsloo}

\affiliation[a]{Departamento de F{\'\i}sica, \\ Universidad de Oviedo,  \\   Avda.~Calvo Sotelo 18, 33007 Oviedo, Spain. \\ \vspace{-0.25cm} } 
\affiliation[b]{The Laboratory for Quantum Gravity and Strings, \\ Department of Mathematics and Applied Mathematics, \\ University of Cape Town, \\ Private Bag, Rondebosch, 7700, South Africa. \\ \vspace{-0.25cm}}
\affiliation[c]{Department of Mathematics, \\ University of Surrey, \\ Guildford, GU2 7XH, United Kingdom.  \\ \vspace{-0.25cm} }

\emailAdd{ylozano@uniovi.es}
\emailAdd{jeff@nassp.uct.ac.za}
\emailAdd{a.prinsloo@surrey.ac.uk}

\abstract{
In this article we extend the construction of giant gravitons from holomorphic surfaces \cite{Mikhailov} to the ABJM correspondence.  We construct a new class of $\tfrac{1}{6}$-BPS M5-branes wrapping 5-manifolds in $S^{7}/\mathbb{Z}_{k}$ and supported by a large angular momentum in the orbifold space.  These orbifold giant gravitons undergo a supersymmetry enhancement to $\tfrac{1}{3}$-BPS and $\tfrac{1}{2}$-BPS configurations in special cases. The compactification of M-theory on $AdS_{4}\times S^{7}/\mathbb{Z}_{k}$ to type IIA superstring theory on $AdS_{4} \times \mathbb{CP}^{3}$ then gives rise to another new class of $\tfrac{1}{6}$-BPS D4 and NS5-branes wrapping 4 and 5-manifolds in $\mathbb{CP}^{3}$.  The D4-branes carry a combination of D0-brane charge and angular momentum in the complex projective space, while the NS5-branes are supported only by D0-brane charge.  Finally, we present a detailed analysis of a one-parameter family of $\tfrac{1}{2}$-BPS M5-brane orbifold giant gravitons, and their D4 and NS5-brane $\mathbb{CP}^{3}$ descendants.
}

\preprint{\begin{flushright}\small{FPAUO--13/06} \\
\small{QGaSLAB--13-03} \\
\small{DMUS--MP--13/13}
\end{flushright}}

\begin{document}
     
\maketitle
\flushbottom

\section{Introduction} \label{section - introduction}

The AdS/CFT correspondence \cite{Maldacena97} provides a nonperturbative definition of string theory (and M-theory) on various anti-de Sitter spacetimes in terms of conformal field theories on flat 
$\mathbb{M}_{d+1}$ spacetimes, which can, in turn, be reformulated by means of the operator-state correspondence on the $\mathbb{R}\times S_{d}$ boundary of $AdS_{d+2}$. Through the AdS/CFT correspondence, the geometry and topology of the bulk spacetime physics must be encoded in the dual field theory. The idea that the macroscopic properties of spacetime are {\it emergent} and not fundamental is neither new nor exclusively string theoretic. Gauge/string theory dualities do, however, provide an excellent laboratory in which to test these ideas.
In particular, through studies of the many brane configurations in string theory and their dual field theoretic descriptions, tremendous progress has been made in understanding the encoding of
\begin{itemize}
\vspace{-0.05cm}
\item{the shape and local position of D-branes in spacetime \cite{Balasubramanian-et-al,CJR,Berenstein-ToyModel, Berenstein-Shape,Berenstein-CollectiveCoords};}
\vspace{-0.05cm}
\item{the supergravity geometries of multiple backreacting D-branes \cite{RdMK-JM-emergent,LLM,RdMK-geometry,RdMK-NI-MS-nontrivial-backgrounds,Ferrari-D3,Ferrari-D3-otherIIB};}
\vspace{-0.05cm}
\item{open strings attached to D-branes, Gauss' law for open string end-points and  \emph{non-planar} integrability in the open string sector \cite{RdMK-strings1,RdMK-strings2,RdMK-strings3,RdMK-Lin,RdMK-giants,BE-OpenSpinChains}}
\vspace{-0.05cm}
\end{itemize}
in the gauge theory. But there is much which still remains to be understood.  Towards this end, as in any laboratory, it is important to have a good sample of specimens to work with.  In this article, we build a catalogue of giant gravitons in the ABJM correspondence \cite{ABJM}: a family of M-branes embedded into $AdS_{4}\times S^{7}/\mathbb{Z}_{k}$ and their NS5/ D-cendents under the compactification to type IIA superstring theory on $AdS_{4}\times \mathbb{CP}^{3}$.
  
Indeed, a natural place to begin studying the emergence of geometry within the framework of the AdS/CFT correspondence is to consider supersymmetric, geometrically non-trivial membranes embedded into supergravity backgrounds with gauge theory duals.  A class of supersymmetric D3-branes in type IIB superstring theory on $AdS_{5} \times S^{5}$, known as giant gravitons in view of their interpretation as large momentum graviton excitations, has proven particularly useful in this context.  The simple example of a $\tfrac{1}{2}$-BPS giant graviton wrapping an $S^{3}$ embedded into $S^{5} \subset \mathbb{C}^{3}$ was originally studied in \cite{MST,GMT} through a Dirac-Born-Infeld (DBI) analysis and later reconstructed\footnote{The holomorphic curve construction of \cite{Mikhailov} extends to a broader class of solutions which include, for example, giant gravitons in the conifold, $T^{1,1}$.  It has also been used to construct `wobbling' dual giant gravitons in $AdS_{5}\subset \mathbb{C}^{2,1}$ \cite{AS-WobblingGiants}.} in \cite{Mikhailov} from holomorphic surfaces in $\mathbb{C}^{3}$, as part of a larger class of D3-brane solutions which includes also $\tfrac{1}{4}$-BPS and $\tfrac{1}{8}$-BPS giant gravitons\footnote{Also of interest is the demonstration of \cite{Kim&Lee} that electromagnetic waves can be introduced on the worldvolumes of these giant gravitons in such a way as to preserve their supersymmetry.}. (Restricted) Schur polynomial operators \cite{CJR,RdMK-strings1,RdMK-strings2}, built from the three complex Higgs fields in the $\mathcal{N}=4$ SYM supermultiplet, form a complete and orthogonal basis of operators dual to (excited) collections of these $\frac{1}{2}$-BPS giant gravitons in $AdS_{5} \times S^{5}$. The operators dual to Mikhailov's $\frac{1}{4}$-BPS and $\tfrac{1}{8}$-BPS giant gravitons embedded into $S^{5}$ are not as yet known, although progress has been made in \cite{PR01} towards the construction of  $\tfrac{1}{8}$-BPS operators in $\mathcal{N}=4$ SYM theory at weak coupling. However, the geometric quantization of the moduli space of $\tfrac{1}{8}$-BPS giant gravitons can be very well described by approximating the holomorphic function by a holomorphic polynomial in $\mathbb{C}^{3}$ whose coefficients are the moduli \cite{Minwalla-et-al,PR}. The Hilbert space obtained in this way gives rise to precisely the partition function over the classical chiral ring of $\tfrac{1}{8}$-BPS states in $\mathcal{N}=4$ SYM theory.

For these statements about the emergence of geometry in the AdS/CFT correspondence to be taken seriously, though, it is important that they not just be statements about AdS$_{5}$/CFT$_{4}$ dualities. More examples are needed, but, fortunately, with the discovery of various AdS$_{4}$/CFT$_{3}$ dualities, more are at hand. The construction of supersymmetric membranes and giant gravitons in the ABJM correspondence \cite{ABJM} has, however, proven considerably more problematic. The ABJM duality links M-theory on $AdS_{4} \times S^{7}/\mathbb{Z}_{k}$ or type IIA superstring theory on $AdS_{4} \times \mathbb{CP}^{3}$ with (two copies of) an $\mathcal{N}=6$ Super-Chern-Simons-matter theory in 2+1 dimensions, with a $U(N)\times U(N)$ gauge group and level numbers $k$ and $-k$, respectively.  This theory has a well-defined 't Hooft coupling $\lambda = \tfrac{N}{k}$, and $k \ll N^{1/5}$ and $k \gg N^{1/5} $ are the M-theory and IIA string theory regimes.  The relation between this M-theory and type IIA superstring theory is well-known \cite{Pope-et-al:1984,Pope-et-al:1997}: the 11D SUGRA geometry $AdS_{4} \times S^{7}/\mathbb{Z}_{k}$ is obtained from $AdS_{4} \times S^{7}$ by an orbifold identification on the fibre of the Hopf fibration
$S^{7} \hookleftarrow  \mathbb{CP}^{3}$.  The compactification of M-theory on $AdS_{4}\times S^{7}/\mathbb{Z}_{k}$ along the Hopf circle to type IIA superstring theory on $AdS_{4}\times \mathbb{CP}^{3}$ can then be implemented as a large $k$ limit of the orbifolding.  Particular examples of 2-branes, 4-branes and 5-branes embedded into $AdS_{4}\times S^{7}/\mathbb{Z}_{k}$ and $AdS_{4} \times \mathbb{CP}^{3}$ were constructed in \cite{NT-giants,HMPS,BT,HLP,GMP,HKY,LP}. A D2-brane giant graviton wrapping an $S^{2}$ embedded into $AdS_{4}$ was constructed in \cite{NT-giants} and studied in more detail in \cite{HMPS}, while various 4-branes and 5-branes (including maximal giant gravitons in $ S^{7}/\mathbb{Z}_{k}$ and $\mathbb{CP}^{3}$) with angular momentum and/or D0-brane charge, were constructed in \cite{HLP}. However, even a simple example of a $\mathbb{CP}^{3}$ giant graviton, being a D4-brane wrapping a 4-manifold of variable size in $\mathbb{CP}^{3}$ and supported by its angular momentum in this complex projective space, proved surprisingly difficult to construct.  Nevertheless, one such solution was eventually found and studied in some detail in \cite{GMP} (and later also in \cite{HKY} in which an independent parameterization of the giant's worldvolume facilitated the computation of various holographic three-point functions). The shape of this object varies with its size and its fluctuation spectrum was shown to exhibit a novel dependence on the size parameter.  This $\mathbb{CP}^{3}$ giant graviton factorizes at maximum size into two D4-branes wrapping non-contractible $\mathbb{CP}^{2} \subset \mathbb{CP}^{3}$ cycles,  which are dual to dibaryon operators in the ABJM model \cite{GLR,MP}. Further discussions of the dual ABJM operators can be found in \cite{SJ,Berenstein-Park,Dey,dMMMP,Caputa:2012,PR02}.   

Beyond these particular examples, however, the general holomorphic surface construction of giant gravitons in $ S^{7}/\mathbb{Z}_{k}$ and $\mathbb{CP}^{3}$ was, until now, unknown for arbitrary $k \in \mathbb{Z}^{+}$. For $k=1$, a class of $\tfrac{1}{8}$-BPS M5-brane giant gravitons embedded into $S^{7}\subset \mathbb{C}^{4}$ was constructed in \cite{Mikhailov} from holomorphic surfaces in the complex manifold $\mathbb{C}^{4}$. These solutions were shown to exhibit a supersymmetry enhancement to $\tfrac{1}{4}$-BPS and $\tfrac{1}{2}$-BPS configurations in special cases. Here we study these \emph{sphere giant gravitons}\footnote{Note that sphere giant gravitons are so-called because they are embedded into the $S^{7}$ compact space rather than in reference to their shape - they are non-spherical objects and some may even be topologically non-trivial.} under the orbifold identification.  In particular, we construct an entirely new class of $\tfrac{1}{6}$-BPS M5-branes embedded into the orbifold compact space $S^{7}/\mathbb{Z}_{k}$. We demonstrate that these {\it orbifold giant gravitons} also enjoy a supersymmetry enhancement to $\tfrac{1}{2}$-BPS and $\tfrac{1}{3}$-BPS configurations in special cases upon which we shall elaborate. A compactification then results in a further new class of $\tfrac{1}{6}$-BPS D4 or NS5-branes embedded into the complex projective space, which we shall call {\it $\mathbb{CP}^{3}$ descendants}. 
The angular momentum along the eleventh fibre direction in the orbifold space gives rise to D0-brane charge.  The D4-branes carry some combination of angular momentum in the complex projective space and D0-brane charge, while the NS5-branes are supported by D0-brane charge only, thus being simply dielectric branes \cite{Myers}.   
A subclass of D4-brane $\mathbb{CP}^{3}$ descendants supported entirely by angular momentum in the complex projective space, genuine \emph{$\mathbb{CP}^{3}$ giant gravitons}, is obtained in the special case in which the holomorphic surface constraint is independent of the eleventh fibre direction.
Now, for the most part, the dual operators in the ABJM model are still unknown, except in the case of these $\mathbb{CP}^{3}$ giant gravitons and at zero coupling \cite{Dey,dMMMP}.  The construction of a more general class of operators dual to orbifold giant gravitons and their $\mathbb{CP}^{3}$ descendants with D0-brane charge must necessarily involve the inclusion of monopole charge, as discussed for $\tfrac{1}{2}$-BPS operators in \cite{SJ,Berenstein-Park}, and remains an important open problem.

The organization of this article is as follows: section \ref{section - sphere giants} opens with a brief review of the construction of $\tfrac{1}{8}$-BPS M5-brane sphere giant gravitons following \cite{Mikhailov}. In the interests of brevity, we omit technical details of the holomorphic curve construction as well as the associated supersymmetry analysis, included rather in appendix \ref{appendix - review}. We then use this construction to discover a new class of $\tfrac{1}{6}$-BPS M5-brane orbifold giant gravitons in section \ref{section - orbifold giants}, and discuss the NS5 and D4-branes to which they descend after a compactification to type IIA string theory on 
$AdS_{4} \times \mathbb{CP}^{3}$.  A brief discussion of the relevant supergravity backgrounds can be found in appendix \ref{appendix - backgrounds}.
We then study the particular example of a one-parameter family of $\tfrac{1}{2}$-BPS orbifold giants in section \ref{section - half-BPS orbifold giants}.  The $\mathbb{CP}^{3}$ descendants of these giant gravitons, either D4 or NS5-branes (depending on the parameter), are constructed in section \ref{section - half-BPS CP3 giants}. The $\tfrac{1}{2}$-BPS NS5-brane solution is identical to the dielectric 5-brane configuration of \cite{HLP} up to a change of worldvolume coordinates. Both the D4 and the NS5-branes pick up an additional coupling to a worldvolume field strength, $\mathcal{F}^{(1)}$, constructed from the RR 1-form potential and therefore
associated with D0-branes `ending' on the D4 or NS5-brane worldvolume.  This happens because the holomorphic function of the orbifold giant graviton ancestor depends on the eleventh fibre direction. In order to correctly describe the D4 and NS5-brane $\mathbb{CP}^{3}$ descendants
we derive a new action with isometric transverse and worldvolume directions, respectively.  This action arises from the M5-brane through a more general reduction ansatz in which the embedding coordinates may depend on the M5-brane worldvolume direction on which the reduction takes place.
Appendix \ref{appendix - actions} contains the details of this construction. We present concluding remarks in section \ref{section - discussion}.

\section{Sphere giant gravitons} \label{section - sphere giants}

\subsection{Sphere giant gravitons from holomorphic surfaces}

It is well-known that a large class of M5-branes, known as sphere giant gravitons, can be embedded into the maximally supersymmetric 11D SUGRA background $AdS_{4}\times S^{7}$. 
The worldvolume $\mathbb{R} \times \Sigma(t)$ was built in \cite{Mikhailov} from a 5-manifold $\Sigma(t) = \mathcal{C}(t)\cap S^{7}$, being the intersection of a holomorphic surface $\mathcal{C}(t)$ in $\mathbb{C}^{4}$ with $S^{7}$.  Here the stationary holomorphic surface $\mathcal{C}$ is defined by 
$f(w_{1},w_{2},w_{3},w_{4}) = 0$ and is brought into motion,
\begin{equation} \label{motion-S7}
\hspace{-0.05cm} \mathcal{C}\hspace{-0.05cm}: \hspace{-0.05cm} f(w_{1},w_{2},w_{3},w_{4}) = 0 \hspace{0.15cm} \longrightarrow \hspace{0.15cm} 
\mathcal{C}(t)\hspace{-0.05cm}: \hspace{-0.05cm} f(w_{1} \, e^{-\frac{i}{2}\dot{\xi} \hspace{0.025cm} t}, w_{2} \, e^{-\frac{i}{2}\dot{\xi} \hspace{0.025cm} t}, w_{3} \, e^{-\frac{i}{2}\dot{\xi} \hspace{0.025cm} t}, w_{4} \, e^{-\frac{i}{2}\dot{\xi} \hspace{0.025cm} t}) = 0,
\end{equation}
by boosting the complex coordinates $w_{a}$ along a preferred direction ${\bf e}^{\parallel}$ (simply the overall phase of the $w_{a}$)
with $\dot{\xi} = \pm 1$ related to the angular velocity. The direction of motion ${\bf e}^{\phi}$ of the giant graviton is then the component of the preferred direction ${\bf e}^{\parallel}$ orthogonal to $\mathrm{T} \Sigma$. 
This preferred direction ${\bf e}^{\parallel} = \textrm{I} \, {\bf e}^{\perp}$ in $\mathrm{T}S^{7}$ is induced by the complex structure of $\mathbb{C}^{4}$ acting on the unit vector ${\bf e}^{\perp}$ in $\mathrm{T}\mathbb{C}^{4}$ which is orthogonal to $\mathrm{T}S^{7}$.  
The complex structure $\textrm{I}$ is, in turn, fixed by our initial choice of coordinates $w_{a} = \rho_{a} \, e^{i\psi_{a}}$ for the complex manifold $\mathbb{C}^{4}$, which breaks the $SO(8)$ symmetry of $S^{7}\subset \mathbb{C}^{4}$ down to $SU(4)$. 
These M5-brane giant gravitons have rigid, rotating geometries which depend on the parameters in the holomorphic function.  This holomorphic surface construction is described in more detail in appendix \ref{appendix - review - surfaces}.

This class of M5-brane sphere giant gravitons was shown in \cite{Mikhailov} to be $\frac{1}{8}$-BPS (in special cases there is a supersymmetry enhancement to $\tfrac{1}{4}$-BPS and $\tfrac{1}{2}$-BPS configurations) with finite energy $H = P_{\xi}$ satisfying the usual BPS bound.  The supersymmetry analysis relies upon an embedding of $AdS_{4}\times S^{7}$ into the flat spacetime  $\mathbb{R}^{2+3} \times \mathbb{C}^{4}$ and the projection of a 32-component Majorana spinor $\Psi$ out of a 64-component complex spinor $\Psi_{+}$.  A review is provided in appendix \ref{appendix - review - susy}. The Killing-Spinor Equations (KSEs) in the flat spacetime $\mathbb{R}^{2+3} \times \mathbb{C}^{4}$ are simply $D_{\mu} \Psi_{+}  = 0$ and the covariantly constant spinor solutions take the form 
\begin{equation}
\Psi_{+} = \mathcal{M}_{\mathbb{R}^{2+3}} \hspace{0.05cm}  \mathcal{M}^{w}_{\mathbb{C}^{4}} \hspace{0.15cm} \Psi_{+}^{0} \equiv \mathcal{M}^{w} \hspace{0.15cm} \Psi_{+}^{0},
\end{equation}
with $\mathcal{M}_{\mathbb{R}^{2+3}}$ shown in (\ref{KS-AdS-solution}) and the $\mathbb{C}^{4}$ dependence given by
\begin{eqnarray}
&& \mathcal{M}^{w}_{\mathbb{C}^{4}} = e^{-\frac{1}{2}\hspace{0.025cm} \psi_{1} \hspace{0.05cm} \gamma^{w}_{5}\gamma^{w}_{9}} 
\ e^{-\frac{1}{2} \hspace{0.025cm} \psi_{2} \hspace{0.05cm} \gamma^{w}_{6}\gamma^{w}_{10}} \
e^{-\frac{1}{2} \hspace{0.025cm} \psi_{3} \hspace{0.05cm} \gamma^{w}_{7}\gamma^{w}_{11}} 
\ e^{-\frac{1}{2}  \hspace{0.025cm} \psi_{4} \hspace{0.05cm} \gamma^{w}_{8}\gamma^{w}_{12}}.
\end{eqnarray} 
Here the flat $\gamma^{w}_{a}$ matrices are associated with the real coordinates $(\rho_{a},\psi_{a})$, being the radii and phases of the complex coordinates $w_{a} = \rho_{a} \,e^{i \psi_{a}}$.  We also make use of flat $\gamma_{0},\ldots, \gamma_{4}$ matrices corresponding to the $\mathbb{R}^{2+3}$ coordinates $(t,r,\theta,\varphi,\hat{R})$ with $\hat{\gamma} \equiv \gamma_{0}\gamma_{1}\gamma_{2}\gamma_{3}$ as in appendix \ref{appendix - review - susy}.
The 64-component complex constant spinor $\Psi_{+}^{0}$ satisfies the conditions
\begin{equation}   \label{conditions-main-text}
\hspace{-0.2cm}
\gamma_{0}\hat{\gamma} \hspace{0.075cm} \Psi_{+}^{0} \equiv  -i \hspace{0.05cm} \dot{\xi} \, \Psi_{+}^{0} \hspace{1.2cm}
 (\gamma^{w}_{5}\gamma^{w}_{9}) (\gamma^{w}_{6} \gamma^{w}_{10}) (\gamma^{w}_{7} \gamma^{w}_{11}) (\gamma^{w}_{8} \gamma^{w}_{12}) \hspace{0.075cm} \Psi^{0}_{-} \equiv \Psi_{+}^{0} \hspace{0.25cm}
\end{equation}
and hence encodes 32 real degrees of freedom.  We can partially label the spinors $\Psi_{+}^{0}$ by the eigenvalues $s^{w}_{i} = \pm 1$ of the three Dirac bilinears $\gamma^{w}_{5}\gamma^{w}_{9}$, $\gamma^{w}_{6}\gamma^{w}_{10}$ and $\gamma^{w}_{7}\gamma^{w}_{11}$:
\begin{equation} 
\hspace{-0.3cm}
\gamma^{w}_{5} \gamma^{w}_{9} \hspace{0.1cm}\Psi_{+}^{0} = is^{w}_{1} \hspace{0.05cm} \Psi_{+}^{0} \hspace{0.9cm}
\gamma^{w}_{6}\gamma^{w}_{10} \hspace{0.1cm} \Psi_{+}^{0} = is^{w}_{2} \hspace{0.05cm} \Psi_{+}^{0} \hspace{0.9cm}
\gamma^{w}_{7}\gamma^{w}_{11} \hspace{0.1cm} \Psi_{+}^{0} = is^{w}_{3} \hspace{0.05cm} \Psi_{+}^{0},
\hspace{0.1cm}
\end{equation}
from which it automatically follows that
\begin{equation}
\gamma^{w}_{8}\gamma^{w}_{12} \hspace{0.1cm} \Psi_{+}^{0} = i \, (s^{w}_{1} \, s^{w}_{2} \, s^{w}_{3}) \hspace{0.075cm} \Psi_{+}^{0},
\end{equation} 
where each of the 8 partial labels $(s^{w}_{1}, s^{w}_{2}, s^{w}_{3}) = (\pm,\pm,\pm)^{w}$ encodes 4 real degrees of freedom.  The $w$ superscripts make explicit our choice of coordinates $w_{a}$, and hence our choice of complex structure for $\mathbb{C}^{4}$ and preferred direction ${\bf e}^{\parallel}$ in $S^{7}$.

The solution of the $\mathbb{R}^{2+3}\times \mathbb{C}^{4}$ background KSEs becomes
\begin{equation} \label{KS-solution-13D-pullback}
\Psi_{+} =  e^{-\frac{1}{2} \hspace{0.025cm} t \hspace{0.05cm} \gamma_{0} \hat{\gamma} } \  
e^{-\frac{1}{2} \hspace{0.025cm} \psi_{1} \hspace{0.05cm} \gamma^{w}_{5}\gamma^{w}_{9} } \ 
e^{-\frac{1}{2} \hspace{0.025cm} \psi_{2} \hspace{0.05cm} \gamma^{w}_{6}\gamma^{w}_{10}} \
e^{-\frac{1}{2} \hspace{0.025cm} \psi_{3} \hspace{0.05cm} \gamma^{w}_{7}\gamma^{w}_{11}} \ 
e^{-\frac{1}{2} \hspace{0.025cm} \psi_{4} \hspace{0.05cm} \gamma^{w}_{8}\gamma^{w}_{12}} \hspace{0.15cm} \Psi_{+}^{0},
\end{equation}
when pulled-back to  $\mathbb{R} \times \mathcal{C}(t)$.  
The kappa symmetry conditions on the pull-back of the spinor $\Psi$ to the worldvolume $\mathbb{R} \times \Sigma(t)$ of the M5-brane giant graviton, associated with the holomorphic surface $\mathcal{C}$ defined by $f(w_{1},w_{2},w_{3},w_{4})=0$, are satisfied if we impose the conditions:
\begin{equation}
\Gamma_{\bar{w}_{a}} \hspace{0.05cm} \Psi_{+}  = 0, \hspace{0.5cm} \text{for all} \hspace{0.2cm} w_{a} \hspace{0.2cm} \text{such that} \hspace{0.2cm} (\p_{w_{a}}f) \neq 0,
\end{equation}    
on the pullback of $\Psi_{+}$ to  $\mathbb{R} \times \mathcal{C}(t)$, the lift of the worldvolume to $\mathbb{R}^{2+3} \times \mathbb{C}^{4}$.  These are additional constraints on (\ref{KS-solution-13D-pullback}), which can be rephrased as
\begin{eqnarray}
&& \hspace{-0.5cm} \Gamma_{\bar{w}_{1}} \Psi_{+} = 0  
\hspace{0.5cm} \Longleftrightarrow \hspace{0.5cm} 
\gamma^{w}_{5}\gamma^{w}_{9} \hspace{0.1cm} \Psi_{+}^{0} = i \hspace{0.075cm} \Psi_{+}^{0}  \hspace{0.76cm} \text{if} \hspace{0.35cm} (\p_{w_{1}}f) \neq 0 \hspace{1.0cm} \\
&& \hspace{-0.5cm} \Gamma_{\bar{w}_{2}} \Psi_{+} = 0  
\hspace{0.5cm} \Longleftrightarrow \hspace{0.5cm} 
\gamma^{w}_{6}\gamma^{w}_{10} \hspace{0.1cm} \Psi_{+}^{0} = i \hspace{0.075cm} \Psi_{+}^{0}  \hspace{0.71cm} \text{if} \hspace{0.35cm} (\p_{w_{2}}f) \neq 0 \hspace{1.0cm} \\
&& \hspace{-0.5cm} \Gamma_{\bar{w}_{3}} \Psi_{+} = 0  
\hspace{0.5cm} \Longleftrightarrow \hspace{0.5cm} 
\gamma^{w}_{7}\gamma^{w}_{11} \hspace{0.1cm} \Psi_{+}^{0} = i \hspace{0.075cm} \Psi_{+}^{0}  \hspace{0.72cm} \text{if} \hspace{0.35cm} (\p_{w_{3}}f) \neq 0 \hspace{1.0cm} \\
&& \hspace{-0.5cm} \Gamma_{\bar{w}_{4}} \Psi_{+} = 0  
\hspace{0.5cm} \Longleftrightarrow \hspace{0.5cm} 
\gamma^{w}_{8}\gamma^{w}_{12} \hspace{0.1cm} \Psi_{+}^{0} = i \hspace{0.075cm} \Psi_{+}^{0}  \hspace{0.73cm} \text{if} \hspace{0.35cm} (\p_{w_{4}}f) \neq 0, \hspace{1.0cm} 
\end{eqnarray}
corresponding to selecting the $s^{w}_{i} = +1$ eigenvalues.  For convenience we shall denote $s^{w}_{4} = s^{w}_{1}s^{w}_{2}s^{w}_{3}$ the eigenvalue of $\gamma^{w}_{8}\gamma^{w}_{12}$ (which is \emph{not} one of our labels).  If the first three conditions are satisfied, the last condition is also automatically satisfied from the second of (\ref{conditions-main-text}).

The classification of these sphere giant gravitons as $\tfrac{1}{2}$-BPS, $\tfrac{1}{4}$-BPS or $\tfrac{1}{8}$-BPS M5-branes therefore simply involves whether our holomorphic surface $\mathcal{C}$ can be written in terms of a function of one, two, three or four complex coordinates $w_{a}$.
In each of these cases, the labels $(s^{w}_{1},s^{w}_{2},s^{w}_{3})$ associated with spinor solutions satisfying the kappa symmetry conditions are shown below:
\begin{equation}
\boxed{
\begin{array} {rcl}
\nonumber f(w_{1}):  & \hspace{0.175cm} (+ \pm \pm)^{w} \hspace{0.5cm} \text{16 of 32 spinors} & \hspace{0.25cm} \Longrightarrow \hspace{0.25cm} \tfrac{1}{2}\text{-BPS} \\
\nonumber 
f(w_{1},w_{2}):  & (+ + \pm)^{w} \hspace{0.5cm} \text{8 of 32 spinors} & \hspace{0.25cm} \Longrightarrow \hspace{0.25cm} \tfrac{1}{4}\text{-BPS} \\
\nonumber 
f(w_{1},w_{2},w_{3}): & (+ + +)^{w} \hspace{0.5cm} \text{4 of 32 spinors} & \hspace{0.25cm} \Longrightarrow \hspace{0.25cm} \tfrac{1}{8}\text{-BPS} \\ 
\nonumber 
f(w_{1},w_{2},w_{3},w_{4}): & (+ + +)^{w} \hspace{0.5cm} \text{4 of 32 spinors} & \hspace{0.25cm} \Longrightarrow \hspace{0.25cm} \tfrac{1}{8}\text{-BPS}
\end{array}
}
\end{equation}
These holomorphic surfaces $\mathcal{C}$ have an evident $U(4)$ symmetry associated with transformations $w_{a} \rightarrow U_{ab} \, w_{b}$ preserving both the $\mathbb{C}^{4}$ metric and the fixed complex structure. 
The M5-brane sphere giant gravitons thus have an evident $SU(4)$ symmetry after the intersection with $S^{7}$. However, we might initially have chosen any set of complex coordinates $w_{a}$ and complex structure $\textrm{I}$ for $\mathbb{C}^{4}$, and so any preferred direction ${\bf e}^{\parallel}$ in $S^{7}$. The full symmetry group is still $SO(8)$, although it is not immediately apparent from this construction.

\subsection{Sphere giant gravitons as the zeros of holomorphic polynomials}

It is possible to approximate any holomorphic function $f(w_{1},w_{2},w_{3},w_{4})$ to arbitrary accuracy by a holomorphic polynomial of degree $n$, if $n$ can be made arbitrarily large.
An $\tfrac{1}{8}$-BPS sphere giant graviton, when boosted into motion, can hence be approximated by the intersection of $S^{7}$ with solutions of
\begin{equation}
\sum_{\ell=1}^{n} \sum_{^{\hspace{0.5cm} n_{1},n_{2},n_{3},n_{4}}_{ \hspace{0.2cm} n_{1}+n_{2}+n_{3}+n_{4}=\ell}} 
c_{n_{1} \hspace{0.025cm} n_{2} \hspace{0.025cm} n_{3} \hspace{0.025cm} n_{4}} \hspace{0.1cm} 
e^{-\frac{i}{2}(n_{1}+n_{2}+n_{3}+n_{4}) \hspace{0.05cm} \dot{\xi} \hspace{0.025cm} t}
\left(w_{1}\right)^{n_{1}} \left(w_{2}\right)^{n_{2}} \left(w_{3}\right)^{n_{3}} \left(w_{4}\right)^{n_{4}} = 0. 
\end{equation}
The coefficients $c_{n_{1}n_{2}n_{3}n_{4}}$ are only defined up to an overall complex rescaling and are therefore elements of a complex projective space $\mathbb{CP}^{n_{\mathcal{C}}}$ with $n_{\mathcal{C}}$ the 
number of possible 4-tuples $(n_{1},n_{2},n_{3},n_{4})$ with $n_{1}+n_{2}+n_{3}+n_{4} \leq n$, being the number of terms in the polynomial.
Note that polynomials with the same intersection with $S^{7}$ result in the same M5-brane giant graviton, which leads to an identification of polynomials which differ
by factors with trivial intersections with $S^{7}$.  Similar observations in $AdS_{5} \times S^{5}$ allowed \cite{Minwalla-et-al,PR} to give a detailed description of the geometric quantization of the phase space of a class of $\tfrac{1}{8}$-BPS D3-brane sphere giant gravitons embedded into this type IIB 10D SUGRA background.

\section{Orbifold giant gravitons} \label{section - orbifold giants}

\subsection{Orbifold giant gravitons from holomorphic surfaces}

Let us consider the 11D SUGRA background $AdS_{4} \times S^{7}/\mathbb{Z}_{k}$ described in appendix \ref{appendix - backgrounds}.  Here $S^{7}$ has been  written as a Hopf fibration over $\mathbb{CP}^{3}$, with 
$S^{7}/\mathbb{Z}_{k}$ the result of an orbifold identification on the Hopf fibre.  In the construction of giant gravitons from holomorphic surfaces,
it is clear that the orientation of the Hopf fibre direction with respect to the preferred direction now becomes important.  We keep $w_{a} = \rho_{a} \, e^{i\psi_{a}}$ as the complex coordinates of $\mathbb{C}^{4}$ associated with our giant graviton construction.  But we shall now define the new complex coordinates $z_{a} = r_{a} \, e^{i\chi_{a}}$ for the orbifold identification, with the Hopf fibre $\tau$ the overall phase of the $z_{a}$.  In this section we show that the choice $(z_{1},z_{2},\bar{z}_{3},\bar{z}_{4}) \equiv (w_{1},w_{2},w_{3},w_{4})$ of complex coordinates (defined up to $U(4)$ symmetry transformations) results in a new class of $\tfrac{1}{6}$-BPS M5-branes under the orbifold identification, which we shall call orbifold giant gravitons, with no configuration losing all its supersymmetry.  Indeed, there is a supersymmetry enhancement to $\tfrac{1}{2}$-BPS and $\frac{1}{3}$-BPS configurations for certain simple holomorphic surfaces.

The holomorphic surface $f(w_{1},w_{2},w_{3},w_{4}) = 0$ in $\mathbb{C}^{4}$ becomes $f(z_{1},z_{2},\bar{z}_{3},\bar{z}_{4}) = 0$ under this coordinate change and is boosted into motion, as before, by taking
\begin{equation} \label{motion-orbifold}
 \mathcal{C}: f(z_{1},z_{2},\bar{z}_{3},\bar{z}_{4}) = 0 \hspace{0.25cm} \longrightarrow \hspace{0.2cm} 
\mathcal{C}(t): f(z_{1} \, e^{-\frac{i}{2}\dot{\xi} \hspace{0.025cm} t},z_{2} \, e^{-\frac{i}{2}\dot{\xi} \hspace{0.02cm} t}, \bar{z}_{3} \, e^{-\frac{i}{2}\dot{\xi} \hspace{0.025cm} t}, \bar{z}_{4} \, e^{-\frac{i}{2}\dot{\xi} \hspace{0.025cm} t}) = 0,
\end{equation}
with $\dot{\xi} = \pm 1$.  The orbifold identification $z_{a} \sim z_{a} \hspace{0.1cm} e^{\frac{2\pi i}{k}}$ or, equivalently, $\tau \sim \tau + \tfrac{2\pi}{k}$ results in the time-dependent surface $\mathcal{C}(t)/\mathbb{Z}_{k}$.   Here   $\tilde{\tau} \equiv \frac{\tau}{k}$ is the orbifold fibre which has the usual $2\pi$ periodicity.  Notice that this new surface in $\mathbb{C}^{4}/\mathbb{Z}_{k}$ may end up wrapped $k$ times on the orbifold fibre $\tilde{\tau}$, as well as possibly moving along it.  The worldvolume $\mathbb{R} \times \Sigma(t)$ of an M5-brane giant graviton embedded into 
$AdS_{4}\times S^{7}/\mathbb{Z}_{k}$ is built as the intersection  $\Sigma(t) = \mathcal{C}(t)/\mathbb{Z}_{k} \cap S^{7}/\mathbb{Z}_{k}$ of the moving surface $\mathcal{C}(t)/\mathbb{Z}_{k}$ in $\mathbb{C}^{4}/\mathbb{Z}_{k}$ with the orbifold space $S^{7}/\mathbb{Z}_{k}$.

Let us now reconsider our previous analysis of the kappa symmetry conditions on the covariantly constant spinor solution $\Psi_{+}$ of the KSEs in flat $\mathbb{R}^{2+3}\times\mathbb{C}^{4}$ spacetime and determine which spinors are projected out under the orbifold identification to $\mathbb{R}^{2+3}\times\mathbb{C}^{4}/\mathbb{Z}_{k}$.  Here 
\begin{equation}
\Psi_{+} = \mathcal{M}_{\mathbb{R}^{2+3}} \hspace{0.05cm}  \mathcal{M}^{z}_{\mathbb{C}^{4}} \hspace{0.15cm} \Psi_{+}^{0} \equiv \mathcal{M}^{z} \hspace{0.15cm} \Psi_{+}^{0},
\end{equation}
with $\mathcal{M}_{\mathbb{R}^{2+3}}$ again given by (\ref{KS-AdS-solution}) and
\begin{eqnarray}
&& \hspace{-0.25cm} \mathcal{M}^{z}_{\mathbb{C}^{4}} = e^{-\frac{1}{2} \hspace{0.025cm} \chi_{1} \hspace{0.05cm} \gamma^{z}_{5}\gamma^{z}_{9} } \ e^{-\frac{1}{2} \hspace{0.025cm} \chi_{2} \hspace{0.05cm} \gamma^{z}_{6}\gamma^{z}_{10} } \
e^{-\frac{1}{2} \hspace{0.025cm} \chi_{3} \hspace{0.05cm} \gamma^{z}_{7}\gamma^{z}_{11} } \ 
e^{-\frac{1}{2} \hspace{0.025cm} \chi_{4} \hspace{0.05cm} \gamma^{z}_{8}\gamma^{z}_{12} } \\
&& \hspace{-0.25cm} \hspace{0.98cm} = e^{-\frac{1}{2} \hspace{0.025cm} \tau \hspace{0.05cm} ( \gamma^{z}_{5}\gamma^{z}_{9} + \gamma^{z}_{6}\gamma^{z}_{10} + \gamma^{z}_{7}\gamma^{z}_{11} + \gamma^{z}_{8}\gamma^{z}_{12}) } \ e^{-\frac{1}{2} \hspace{0.025cm} \varphi_{1} \hspace{0.05cm} \gamma^{z}_{6}\gamma^{z}_{10} } \
e^{\frac{1}{2} \hspace{0.025cm} \chi \hspace{0.05cm} (\gamma^{z}_{7}\gamma^{z}_{11} + \gamma^{z}_{8}\gamma^{z}_{12}) } \ 
e^{\frac{1}{2}  \hspace{0.025cm} \varphi_{2} \hspace{0.05cm} \gamma^{z}_{8}\gamma^{z}_{12}}.
\end{eqnarray} 
We make use of the real coordinates $(r_{a},\chi_{a})$, the radii and phases of the $z_{a} = r_{a} \, e^{i\chi_{a}}$ complex coordinates, and then change to the new angular coordinates $(\tau,\chi,\varphi_{1},\varphi_{2})$ discussed in appendix \ref{appendix - backgrounds}. We can again partially label the constant spinors $\Psi_{+}^{0}$ by the eigenvalues $s^{z}_{a}$ of the Dirac bilinears $\gamma^{z}_{5}\gamma^{z}_{9}$, $\gamma^{z}_{6}\gamma^{z}_{10}$ and $\gamma^{z}_{7}\gamma^{z}_{11}$:
\begin{equation}
\gamma^{z}_{5}\gamma^{z}_{9} \hspace{0.1cm}\Psi_{+}^{0} = is^{z}_{1} \hspace{0.1cm} \Psi_{+}^{0} \hspace{1.0cm}
\gamma^{z}_{6}\gamma^{z}_{10} \hspace{0.1cm} \Psi_{+}^{0} = is^{z}_{2} \hspace{0.1cm} \Psi_{+}^{0} \hspace{1.0cm}
\gamma^{z}_{7}\gamma^{z}_{11} \hspace{0.1cm} \Psi_{+}^{0} = is^{z}_{3} \hspace{0.1cm} \Psi_{+}^{0},
\hspace{0.1cm}
\end{equation}
from which it automatically follows that
\begin{equation}
\gamma^{z}_{8}\gamma^{z}_{12} \hspace{0.1cm} \Psi_{+}^{0} = i \hspace{0.05cm} (s^{z}_{1} \, s^{z}_{2} \, s^{z}_{3}) \hspace{0.1cm} \Psi_{+}^{0}.
\end{equation}
Recall that each of these partial labels is associated with 4 spinor degrees of freedom. Now let us make the orbifold identification $\tau \sim \tau + \tfrac{2\pi}{k}$ on the phases of the $z_{a}$. The spinor solutions $ \Psi_{+}$ contain the explicit dependence
\begin{eqnarray} 
&& \mathcal{M}^{z}  \sim
e^{-\frac{1}{2} \hspace{0.025cm} \tau \hspace{0.05cm} 
(\gamma^{z}_{5}\gamma^{z}_{9} + \gamma^{z}_{6}\gamma^{z}_{10} + \gamma^{z}_{7}\gamma^{z}_{11} + \gamma^{z}_{8}\gamma^{z}_{12} )} 
= e^{-\frac{1}{2} \hspace{0.025cm} \frac{\tilde{\tau}}{k} \hspace{0.05cm} (\gamma^{z}_{5}\gamma^{z}_{9} + \gamma^{z}_{6}\gamma^{z}_{10} + \gamma^{z}_{7}\gamma^{z}_{11} + \gamma^{z}_{8}\gamma^{z}_{12} )},   \hspace{0.5cm} 
\end{eqnarray}
on the fibre direction and therefore do not satisfy the usual periodic boundary conditions (up to a sign) when $s^{z}_{1} + s^{z}_{2} + s^{z}_{3} + s^{z}_{1} \, s^{z}_{2} \, s^{z}_{3} \neq 0$.  Thus, as argued in \cite{NT-giants}, the spinors associated with the labels $(+ + +)^{z}$ and $(- - -)^{z}$ do not survive the orbifolding.  By retaining only the other 6 of 8 partial labels, we ensure the Killing spinors are well-defined on 
$AdS_{4}\times \mathbb{C}^{4}/\mathbb{Z}_{k}$.  These spinors $\Psi_{+}$ can be projected onto solutions $\Psi$ of the $AdS_{4}\times S^{7}/\mathbb{Z}_{k}$ KSEs using the projection described in appendix \ref{appendix - review - susy}. It thus follows that the 11D SUGRA background $AdS_{4}\times S^{7}/\mathbb{Z}_{k}$ retains 24 of the original 32 supersymmetries.

The pullback of these 24 covariantly constant spinors $\Psi_{+}$ to $\mathbb{R}\times \mathcal{C}(t)/\mathbb{Z}_{k}$ is
\begin{eqnarray} \label{spinor-z-pullback}
&& \hspace{-0.25cm} \Psi_{+}
 =  e^{-\frac{1}{2} \hspace{0.025cm} t \hspace{0.05cm} \gamma_{0} \hat{\gamma} } \  
e^{-\frac{1}{2} \hspace{0.025cm} \chi_{1} \hspace{0.05cm} \gamma^{z}_{5}\gamma^{z}_{9} } \ 
e^{-\frac{1}{2} \hspace{0.025cm} \chi_{2} \hspace{0.05cm} \gamma^{z}_{6}\gamma^{z}_{10} } \
e^{-\frac{1}{2} \hspace{0.025cm} \chi_{3} \hspace{0.05cm} \gamma^{z}_{7}\gamma^{z}_{11} } \ 
e^{-\frac{1}{2} \hspace{0.025cm} \chi_{4} \hspace{0.05cm} \gamma^{z}_{8}\gamma^{z}_{12} } \hspace{0.15cm} \Psi_{+}^{0}  \hspace{0.5cm} \\
&& \hspace{-0.25cm} \hspace{0.68cm} =  e^{-\frac{1}{2} \hspace{0.025cm} t \hspace{0.05cm} \gamma_{0} \hat{\gamma} } \  
e^{-\frac{1}{2} \hspace{0.025cm} \varphi_{1} \hspace{0.05cm} \gamma^{z}_{6}\gamma^{z}_{10} } \
e^{\frac{1}{2} \hspace{0.025cm} \chi \hspace{0.05cm} (\gamma^{z}_{7}\gamma^{z}_{11} + \gamma^{z}_{8}\gamma^{z}_{12}) } \ 
e^{\frac{1}{2} \hspace{0.025cm} \varphi_{2} \hspace{0.05cm} \gamma^{z}_{8}\gamma^{z}_{12} } \hspace{0.15cm} \Psi_{+}^{0} 
\label{pullback-KS-orbifold},
\end{eqnarray} 
with the $\chi_{a}$ phases (or the $\chi$ and $\varphi_{i}$ phases) constrained to $\mathcal{C}(t)/\mathbb{Z}_{k}$ and hence to the worldvolume of the orbifold giant graviton.  
Although there is no explicit dependence on $\tilde{\tau}$, due to our removal of the labels $(+ + +)^{z}$ and $(- - - )^{z}$, there may be an implicit $\tilde{\tau}$-dependence through the phases $\chi$ and $\varphi_{i}$.
Under the orbifold identification, it is hence possible for the pullback of the spinor (\ref{pullback-KS-orbifold}) to pick-up non-periodic boundary conditions, with repetition (up to a sign) only upon traversing the $\tilde{\tau}$ circle $k$ times.  This is not inconsistent, however, if we interpret it as an indication that the M5-brane has become wrapped $k$ times on the Hopf fibre direction $\tilde{\tau}$, leading to multi-valued Killing spinors on its worldvolume.

The kappa symmetry conditions on $\Psi$ are satisfied if
\begin{eqnarray}
&& \hspace{-0.5cm} \Gamma_{\bar{z}_{1}} \Psi_{+} = 0  
\hspace{0.5cm} \Longleftrightarrow \hspace{0.5cm} 
\gamma^{z}_{5}\gamma^{z}_{9} \hspace{0.1cm} \Psi_{+}^{0} = i \hspace{0.075cm} \Psi_{+}^{0}  \hspace{1.18cm} \text{if} \hspace{0.35cm} (\p_{z_{1}}f) \neq 0 \hspace{1.0cm} \\
&& \hspace{-0.5cm} \Gamma_{\bar{z}_{2}} \Psi_{+} = 0  
\hspace{0.5cm} \Longleftrightarrow \hspace{0.5cm} 
\gamma^{z}_{6}\gamma^{z}_{10} \hspace{0.1cm} \Psi_{+}^{0} = i \hspace{0.075cm} \Psi_{+}^{0}  \hspace{1.04cm} \text{if} \hspace{0.35cm} (\p_{z_{2}}f) \neq 0 \hspace{1.0cm} \\
&& \hspace{-0.5cm} \Gamma_{z_{3}} \Psi_{+} = 0  
\hspace{0.5cm} \Longleftrightarrow \hspace{0.5cm} 
\gamma^{z}_{7}\gamma^{z}_{11} \hspace{0.1cm} \Psi_{+}^{0} = -i \hspace{0.075cm} \Psi_{+}^{0}  \hspace{0.72cm} \text{if} \hspace{0.35cm} (\p_{\bar{z}_{3}}f) \neq 0 \hspace{1.0cm} \\
&& \hspace{-0.5cm} \Gamma_{z_{4}} \Psi_{+} = 0  
\hspace{0.5cm} \Longleftrightarrow \hspace{0.5cm} 
\gamma^{z}_{8}\gamma^{z}_{12} \hspace{0.1cm} \Psi_{+}^{0} = -i \hspace{0.075cm} \Psi_{+}^{0}  \hspace{0.73cm} \text{if} \hspace{0.35cm} (\p_{\bar{z}_{4}}f) \neq 0. \hspace{1.0cm} 
\end{eqnarray}
Alternatively, simply note that $(\pm,\pm,\pm)^{w} = (\pm,\pm,\mp)^{z}$, so we can make use of the original kappa symmetry conditions and project out the partial labels $(+ + -)^{w}$ and $(- - +)^{w}$. We obtain the following classification:
\begin{equation}
\boxed{
\begin{array}{rcl}
f(z_{1}):  & \hspace{0.15cm} (+ + -)^{z} \hspace{0.15cm} (+ - +)^{z} \hspace{0.15cm} (+ - -)^{z} \hspace{0.35cm} 
\text{12 of 24 spinors} & \hspace{0.25cm} \Longrightarrow \hspace{0.25cm} \tfrac{1}{2}\text{-BPS} \\
f(\bar{z}_{3}):  & \hspace{0.2cm} (- - +)^{z} \hspace{0.15cm} (- + -)^{z} \hspace{0.15cm} (- + +)^{z} \hspace{0.35cm} 
\text{12 of 24 spinors} & \hspace{0.25cm} \Longrightarrow \hspace{0.25cm} \tfrac{1}{2}\text{-BPS} \hspace{0.5cm} \\
&& \nonumber \\
f(z_{1},\bar{z}_{3}):  & (+ + -)^{z} \hspace{0.15cm} (+ - -)^{z} \hspace{2.23cm} \text{8 of 24 spinors} 
& \hspace{0.25cm} \Longrightarrow \hspace{0.25cm} \tfrac{1}{3}\text{-BPS} \\
\hspace{-0.1cm}
f(z_{1},z_{2}):  & (+ + -)^{z} \hspace{4.03cm} \text{4 of 24 spinors} & \hspace{0.25cm} \Longrightarrow \hspace{0.25cm} \tfrac{1}{6}\text{-BPS} \\
f(\bar{z}_{3},\bar{z}_{4}):  & (+ + -)^{z} \hspace{4.03cm} \text{4 of 24 spinors} & \hspace{0.25cm} \Longrightarrow \hspace{0.25cm} \tfrac{1}{6}\text{-BPS} \\
&& \nonumber \\
f(z_{1},z_{2},\bar{z}_{3}): & (+ + -)^{z} \hspace{4.03cm} \text{4 of 24 spinors} & \hspace{0.25cm} \Longrightarrow \hspace{0.25cm} \tfrac{1}{6}\text{-BPS} \\
f(z_{1},\bar{z}_{3},\bar{z}_{4}): & (+ + -)^{z} \hspace{4.03cm} \text{4 of 24 spinors} & \hspace{0.25cm} 
\Longrightarrow \hspace{0.25cm} \tfrac{1}{6}\text{-BPS} \\
&& \nonumber \\
\hspace{-0.1cm}
f(z_{1},z_{2},\bar{z}_{3},\bar{z}_{4}): & (+ + -)^{z} \hspace{4.03cm} \text{4 of 24 spinors} & \hspace{0.25cm} \Longrightarrow \hspace{0.25cm} \tfrac{1}{6}\text{-BPS} 
\end{array}
}
\end{equation}

This construction has a $U(4)$ symmetry associated with transformations preserving both the metric and complex structure of $\mathbb{C}^{4}$ before the orbifold reduction:
\begin{equation}
(z_{1},z_{2},\bar{z}_{3},\bar{z}_{4})^{T} \hspace{0.1cm} \rightarrow \hspace{0.1cm} U (z_{1},z_{2},\bar{z}_{3},\bar{z}_{4})^{T}, \hspace{0.75cm}
\text{with} \hspace{0.5cm} U \in U(4),
\end{equation}
becoming an $SU(4)$ symmetry when the surfaces $\mathcal{C}(t)/\mathbb{Z}_{k}$ are intersected with $S^{7}/\mathbb{Z}_{k}$.

\subsection{Orbifold giant gravitons as the zeros of holomorphic polynomials}

The holomorphic function $f(w_{1},w_{2},w_{3},w_{4})$, rewritten in terms of the $z_{a}$ coordinates, is given by $f(z_{1},z_{2},\bar{z}_{3},\bar{z}_{4})$ and can again be approximated to arbitrary accuracy by a polynomial of degree $n$, if $n$ can be made arbitrarily large.
The surface $\mathcal{C}(t)$ in $\mathbb{C}^{4}$, when boosted into motion, can be approximated by the solution of
\begin{eqnarray}
&&  \sum_{\ell=1}^{n} \sum_{^{\hspace{0.5cm} n_{1},n_{2},n_{3},n_{4}}_{ \hspace{0.2cm} n_{1}+n_{2}+n_{3}+n_{4}=\ell}}  
c_{n_{1} \hspace{0.025cm} n_{2} \hspace{0.025cm} n_{3} \hspace{0.025cm} n_{4}} \hspace{0.1cm} 
e^{-\frac{i}{2}(n_{1}+n_{2}+n_{3}+n_{4}) \hspace{0.05cm} \dot{\xi} \hspace{0.025cm} t}
\left(z_{1}\right)^{n_{1}} \left(z_{2}\right)^{n_{2}} \left(\bar{z}_{3}\right)^{n_{3}} \left(\bar{z}_{4}\right)^{n_{4}} = 0. \hspace{1.0cm}
\end{eqnarray}
Let us now consider the orbifold reduction.  We can rewrite the complex coordinates $z_{a}$ in terms of the radii $r_{a}$ and redefined phases $\tau$, $\chi$ and $\varphi_{i}$. Setting $\tau = \tfrac{\tilde{\tau}}{k}$, where $\tilde{\tau}$ has the usual $2\pi$ periodicity, we then approximate the surface $\mathcal{C}(t)/\mathbb{Z}_{k}$ in $\mathbb{C}^{4}/\mathbb{Z}_{k}$ by
\begin{eqnarray}
&& \hspace{-0.25cm} \sum_{n_{1},n_{2},n_{3},n_{4}} c_{n_{1}n_{2}n_{3}n_{4}} \hspace{0.1cm}  \nonumber e^{-\frac{i}{2}(n_{1}+n_{2}+n_{3}+n_{4}) \hspace{0.05cm} \dot{\xi} \hspace{0.025cm} t} \, \,
(r_{1})^{n_{1}} \, (r_{2})^{n_{2}} \, (r_{3})^{n_{3}} \, (r_{4})^{n_{4}}  \\
&& \hspace{-0.25cm} \hspace{2.8cm} \times \ e^{\frac{1}{2}(n_{1}+n_{2}-n_{3}-n_{4}) \, \frac{\tilde{\tau}}{k}} \ e^{\frac{i}{2}(n_{3}+n_{4})\chi} \ e^{\frac{i}{2}n_{2}\varphi_{1}}  \ e^{\frac{i}{2}n_{4}\varphi_{2}} = 0, \hspace{2.0cm} \label{ansatz - orbifold giants}
\end{eqnarray}
which, when intersected with $S^{7}/\mathbb{Z}_{k}$, gives the (spatial) worldvolume of an M5-brane orbifold giant graviton. This polynomial is clearly dependent on the orbifold fibre, in general.  This allows for the possibility of motion along the eleventh fibre direction $\tilde{\tau}$ and/or, if $\tilde{\tau}$ is a worldvolume direction, the M5-brane embedding coordinates becoming $\tilde{\tau}$-dependent. 

It is only in the special case in which all the coefficients $c_{n_{1}n_{2}n_{3}n_{4}}$ vanish, except those with
$n_{1} + n_{2} = n_{3} + n_{4} \equiv m$, that the holomorphic function is independent of $\tilde{\tau}$.  There is then no motion along the orbifold fibre $\tilde{\tau}$ and no dependence on the worldvolume direction $\tilde{\tau}$ in the M5-brane orbifold giant graviton embedding coordinates. The ansatz is thus considerably more straightforward:
\begin{equation} \label{ansatz - orbifold giants - independent}
\sum_{m=1}^{\frac{n}{2}} \sum_{n_{1}=1}^{m} \sum_{n_{3}=1}^{m} c_{n_{1}n_{2}n_{3}n_{4}} \
e^{-im \hspace{0.025cm} \dot{\xi} \hspace{0.025cm} t} \hspace{0.15cm}  (r_{1})^{n_{1}} \, (r_{2})^{n_{2}} \, (r_{3})^{n_{3}} \, (r_{4})^{n_{4}}  \hspace{0.15cm} 
e^{\frac{i}{2} \,m\chi} \ e^{\frac{i}{2} \, n_{2}\varphi_{1}}  \ e^{\frac{i}{2} \,n_{4}\varphi_{2}} = 0,
\end{equation}
with $n_{2} = m-n_{1}$ and $n_{4}=m-n_{3}$. Here the polynomial depends only on the combinations $z_{1}\bar{z}_{3}$, $z_{1}\bar{z}_{4}$, $z_{2}\bar{z}_{3}$ and $z_{2}\bar{z}_{4}$.  There is an obvious link between this special subclass of $\tfrac{1}{6}$-BPS orbifold giant gravitons and the restricted Schur polynomial operators built from composite scalars $A_{1}B^{\dag}_{1}$, $A_{1}B^{\dag}_{2}$, $A_{2}B^{\dag}_{1}$ and $A_{2}B^{\dag}_{2}$  in the ABJM model, which were constructed in \cite{dMMMP}.  The operators dual to the more general class of $\tfrac{1}{6}$-BPS orbifold giant gravitons associated with the ansatz (\ref{ansatz - orbifold giants}) are not presently known.

\subsection{$\mathbb{CP}^{3}$ descendants} \label{subsection - CP3 giant descendants}

The compactification on the orbifold fibre $\tilde{\tau}$, through which M-theory on $AdS_{4}\times S^{7}/\mathbb{Z}_k$ is reduced to type IIA superstring theory on $AdS_{4}\times \mathbb{CP}^{3}$, can be seen as the limit of large $k$.  This new class of $\tfrac{1}{6}$-BPS M5-brane orbifold giant gravitons therefore gives rise to a new class of $\tfrac{1}{6}$-BPS D4 and NS5-brane descendants, wrapping 4-manifolds and 5-manifolds in the complex projective space, with the same supersymmetry as their M5-brane ancestors. These so-called $\mathbb{CP}^{3}$ descendants are associated with surfaces $f(z_{1},z_{2},\bar{z}_{3},\bar{z}_{4}) = 0$, boosted into motion as in (\ref{motion-orbifold}), but now involving the $\mathbb{CP}^{3}$ homogenous coordinates $z_{a}$, which are parameterized only by the phases $\chi$, $\varphi_{1}$ and $\varphi_{2}$, and \emph{not} by the overall phase $\tau \equiv \tfrac{\tilde{\tau}}{k}$.  We expect the $\mathbb{CP}^{3}$ descendants to pick up an additional D0-brane charge (as in \cite{HLP,LP}), if the function $f$ depends on the eleventh fibre direction $\tilde{\tau}$.  We do, indeed, observe a coupling in the D4 and NS5-brane action of section \ref{section - half-BPS CP3 giants} to a worldvolume field strength, $\mathcal{F}^{(1)}$, associated with D0-branes `ending' on the worldvolume. We present explicit constructions of $\tfrac{1}{2}$-BPS orbifold giant gravitons and their $\mathbb{CP}^{3}$ descendants in sections \ref{section - half-BPS orbifold giants} and \ref{section - half-BPS CP3 giants} to demonstrate this effect.

The special cases of M5-brane orbifold giant gravitons associated with functions of the form $f(z_{1}\bar{z}_{3},z_{1}\bar{z}_{4},z_{2}\bar{z}_{3},z_{2}\bar{z}_{4})$  have ans\"{a}tze which remain unaltered by the orbifold identification and subsequent compactification.  The worldvolume $\mathbb{R} \times \Sigma(t)$ can then be obtained by boosting the surface $\Sigma$ into motion,
\begin{eqnarray}
\nonumber \Sigma:  f(z_{1}\bar{z}_{3},z_{1}\bar{z}_{4},z_{2}\bar{z}_{3},z_{2}\bar{z}_{4})=0 \hspace{0.15cm} \longrightarrow \hspace{0.15cm}
\Sigma(t) : \ f(z_{1}\bar{z}_{3} \, e^{- \hspace{0.025cm}i \, \dot{\xi} \hspace{0.025cm} t}, 
z_{1}\bar{z}_{4} \, e^{- \hspace{0.025cm} i \, \dot{\xi} \hspace{0.025cm} t}, 
z_{2}\bar{z}_{3} \, e^{- \hspace{0.025cm} i \, \dot{\xi} \hspace{0.025cm} t}, 
z_{2}\bar{z}_{4} \, e^{- \hspace{0.025cm} i \, \dot{\xi} \hspace{0.025cm} t}) = 0, \hspace{-0.3cm} \\
\end{eqnarray}
with $\dot{\xi} = \pm 1$.  
There is a supersymmetry enhancement from $\tfrac{1}{6}$-BPS to $\tfrac{1}{3}$-BPS when the function $f(z_{1}\bar{z}_{3})$ depends on only one combination, say $z_{1}\bar{z}_{3}$.
These surfaces $\Sigma(t)$ may be approximated to arbitrary accuracy by solutions of (\ref{ansatz - orbifold giants - independent}), with the $z_{a}$ now the homogenous coordinates of $\mathbb{CP}^{3}$. This special subclass of $\mathbb{CP}^{3}$ descendants, being D4-brane $\mathbb{CP}^{3}$ giant gravitons supported entirely by their angular momentum in the complex projective space, carry no extra D0-brane charge and are described by the standard DBI plus CS action.

One such example is the $\mathbb{CP}^{3}$ giant graviton of \cite{GMP,HKY}, which we now view as the $\mathbb{CP}^{3}$ descendant of the $\tfrac{1}{3}$-BPS orbifold giant graviton associated with the polynomial function 
\begin{equation}
f(z_{1}\bar{z}_{3}) = z_{1}\bar{z}_{3} - (2R)^{2}\sqrt{1-\alpha_{0}^{2}} \, , \hspace{0.5cm}
\end{equation}
with $\alpha_{0} \in [0,1]$ the constant size parameter.  This D4-brane $\mathbb{CP}^{3}$ giant graviton should, at least at zero coupling, be dual to a Schur polynomial operator
$\chi_{R}(A_{1}B_{1}^{\dag})$ constructed from the composite scalar field $A_{1}B^{\dag}_{1}$ and labeled by the totally anti-symmetric representation $R$ of the permutation group \cite{Dey,dMMMP}.

\section{$\tfrac{1}{2}$-BPS orbifold giant gravitons} \label{section - half-BPS orbifold giants}

The simplest $\tfrac{1}{2}$-BPS sphere giant gravitons are associated with linear holomorphic polynomial functions of a single complex coordinate 
\begin{equation}
f(w_{a}) = w_{a} - (2R) \sqrt{1-\alpha_{0}^2} \, ,
\end{equation} with $\alpha_{0} \in [0,1]$ the radial size of the $S^{5} \subset S^{7}$ worldvolume.  After the orbifold identification, the $\tfrac{1}{2}$-BPS orbifold giant gravitons associated with these linear functions will differ, depending on the alignment between their directions of motion and the fibre direction $\tilde{\tau}$.  Ideally, we would like to describe a family of $\tfrac{1}{2}$-BPS orbifold giant gravitons parameterized by the proportion of their motion along $\tau$ before the orbifolding.

Towards this end, let us modify the parameterization of appendix \ref{appendix - backgrounds} by an overall phase, which is dependent on a new parameter $\beta \in [0,1]$:
 \begin{eqnarray}
\nonumber && z_{1} = r_{1} \hspace{0.05cm} e^{i\chi_{1}} 
= (2R) \hspace{0.05cm} \cos{\zeta} \sin{\tfrac{\theta_{1}}{2}} \hspace{0.1cm} e^{i \{\tau - \beta \varphi_{1}\}} \hspace{1.2cm} \\
\nonumber && z_{2} = r_{2} \hspace{0.05cm} e^{i\chi_{2}} 
= (2R) \hspace{0.05cm} \cos{\zeta} \cos{\tfrac{\theta_{1}}{2}} \hspace{0.1cm} e^{i \{\tau + (1-\beta) \varphi_{1}\}} \hspace{1.2cm}  \\ 
\nonumber &&  z_{3} = r_{3} \hspace{0.05cm} e^{i\chi_{3}} 
= (2R) \hspace{0.05cm} \sin{\zeta} \sin{\tfrac{\theta_{2}}{2}} \hspace{0.1cm} e^{i \{\tau - \chi - \beta\varphi_{1}\} } \hspace{1.25cm} \\
&& z_{4} = r_{4} \hspace{0.05cm} e^{i\chi_{4}} 
= (2R) \hspace{0.05cm} \sin{\zeta} \cos{\tfrac{\theta_{2}}{2}} \hspace{0.1cm} e^{i \{\tau - \chi - \varphi_{2} - \beta \, \varphi_{1}\} }, \hspace{1.2cm} 
\end{eqnarray}
with $\tau = \tfrac{\tilde{\tau}}{k}$ after the orbifold identification. This is equivalent to shifting $\tau \rightarrow \tau - \beta \, \varphi_{1}$ and thus does not change the metric of the complex projective space.
Here now
\begin{eqnarray}
&& \hspace{-0.35cm} \mathcal{A}  =  
- \sin^{2}{\zeta}  \hspace{0.15cm} d\chi + (\cos^{2}{\zeta}\cos^{2}{\tfrac{\theta_{1}}{2}} -\beta) \hspace{0.15cm} d\varphi_{1} - \sin^{2}{\zeta} \cos^{2}{\tfrac{\theta_{2}}{2}} \hspace{0.15cm} d\varphi_{2}  \hspace{1.2cm}
\end{eqnarray}
is the only redefinition needed for the background spacetime (see appendix \ref{appendix - backgrounds}).   

We shall take as our holomorphic function 
\begin{equation}
f(z_{2}) = z_{2} - (2R)\sqrt{1-\alpha_{0}^{2}} \, .
\end{equation} 
The moving surface $\mathcal{C}(t)$ is $f(z_{2} \, e^{-\frac{i}{2} \hspace{0.025cm} \dot{\xi} \hspace{0.025cm} t}) = 0$ which, when intersected with $S^{7}$ describes the (spatial) worldvolume of the sphere giant graviton and leads to the expression
\begin{equation} 
\cos{\zeta} \cos{\tfrac{\theta_{1}}{2}} \hspace{0.2cm} e^{i\{\tau + (1-\beta) \varphi_{1}\}} 
= \sqrt{1- \alpha_{0}^{2}} \hspace{0.25cm} \, e^{\frac{i}{2} \hspace{0.025cm} \xi(t)}, 
\hspace{0.5cm} \text{where} \hspace{0.25cm} \xi(t) = \dot{\xi} \, t = \pm \, t,
\end{equation}
with the other phases independent of the worldvolume coordinates. We set $\tau \equiv \tfrac{\tilde{\tau}}{k}$ to obtain the surface  $\mathcal{C}(t)/\mathbb{Z}_{k}$ which must be intersected with $S^{7}/\mathbb{Z}_{k}$.
The worldvolume of the orbifold giant graviton is then described by the radial coordinates $\zeta \in [0,\tfrac{\pi}{2}]$ and $\theta_{i}\in [0,\pi]$, satisfying the constraint
\begin{equation} 
\cos^{2}{\zeta} \cos^{2}{\tfrac{\theta_{1}}{2}} 
= 1- \alpha_{0}^{2}, \hspace{1.0cm}
\end{equation}
and the angular coordinates $\sigma,\chi,\varphi_{2} \in [0,2\pi]$ in terms of which the phases are
\begin{equation} \label{angular-ansatz}
\tau \equiv \tfrac{\tilde{\tau}}{k} = \tfrac{1}{2} \, \beta \, \xi(t) +  \left( 1- \beta\right)\sigma, \hspace{0.5cm}
\varphi_{1} = \tfrac{1}{2} \, \xi(t) - \sigma, \hspace{0.5cm} \chi \hspace{0.5cm} \text{and} \hspace{0.5cm} \varphi_{2}. \hspace{0.5cm}
\end{equation}
This M5-brane configuration is twisted on the Hopf fibre direction for all $\beta \neq 1$:
\begin{equation} \label{twist}
\varphi_{1} = \frac{1}{(1-\beta)} \left\{\tfrac{1}{2} \,  \xi(t) - \tau \right\} 
= \frac{1}{(1-\beta)} \left\{\tfrac{1}{2} \, \xi(t) - \frac{\tilde{\tau}}{k} \right\}
\end{equation}
by which we mean that the embedding coordinate $\varphi_{1}(t,\tilde{\tau})$ depends on the eleventh fibre direction.
Here $\tilde{\tau} \in [0,2\pi k]$ may be regarded as an alternative worldvolume coordinate to $\sigma$, after a coordinate redefinition involving a mixing with the worldvolume time $t$.  The orbifold giant graviton is wrapped on $\tilde{\tau}$ and thus we expect its $\mathbb{CP}^{3}$ descendant (constructed in section \ref{section - half-BPS CP3 giants}) to be a D4-brane giant graviton with D0-brane charge, as well as angular momentum in the complex projective space.

The special case of $\beta = 1$ has
\begin{equation} \label{beta1-case}
\tau \equiv \tfrac{\tilde{\tau}}{k} = \tfrac{1}{2} \, \xi(t) \hspace{0.75cm} \text{and} \hspace{0.75cm} \varphi_{1} = \tfrac{1}{2} \, \xi(t) - \sigma
\end{equation}
with motion along $\tilde{\tau}$ and $\varphi_{1}$, but no wrapping on the fibre direction.  The $\mathbb{CP}^{3}$ descendant must therefore be an NS5-brane supported only by D0-brane charge.

\subsection{$\mathbb{R} \times S^{7}/\mathbb{Z}_{k}$ background} 

We shall parameterize the $\mathbb{R}\times S^{7}/\mathbb{Z}_{k} \subset AdS_{4} \times S^{7}/\mathbb{Z}_k$ subspace, into which the $\tfrac{1}{2}$-BPS orbifold giant is embedded, in terms of the anti-de Sitter time $t$, the angular coordinates $(\tilde{\tau},\chi,\varphi_{1},\varphi_{2})$ and the radial coordinates 
$(\alpha,u,z)$.  Here we define
\begin{equation}  \label{definition-radial-coordinates}
\cos^{2}{\zeta} = (1-\alpha^{2}u), \hspace{0.75cm} 
\cos^{2}{\tfrac{\theta_{1}}{2}} = \frac{(1-\alpha^{2})}{(1-\alpha^{2}u)} \hspace{0.75cm} \text{and} \hspace{0.75cm}
 \cos^{2}{\tfrac{\theta_{2}}{2}} = z.
\end{equation}
The parameter $\alpha_{0} \in [0,1]$ has been promoted to a radial coordinate  orthogonal to the giant's worldvolume, while $u,z \in [0,1]$ are radial worldvolume coordinates.

The metric can be written as
\begin{equation} \hspace{-0.5cm}
ds^{2} = R^{2} \left\{ -dt^{2} + 4 \hspace{0.05cm} ds_{S^{7}/\mathbb{Z}_k}^{2} \right\} = R^{2} \left\{ -dt^{2} + ds_{\textrm{rad}}^{2} + ds_{\textrm{ang}}^{2} \right\},
\end{equation}
where the radial and angular parts are
\begin{eqnarray}
&& \hspace{-0.25cm} ds_{\textrm{rad}}^{2} = \frac{4 \hspace{0.075cm} d\alpha^{2}}{\left( 1-\alpha^{2} \right)} + \frac{\alpha^{2} \hspace{0.05cm} du^{2} }{u\left( 1-u \right)} + \frac{\alpha^{2} u  \hspace{0.1cm} dz^{2}}{z\left(1-z\right)}\\
&& \nonumber \\
&& \hspace{-0.25cm} ds_{\textrm{ang}}^{2} = 4\hspace{0.025cm}  \alpha^{2}  \left( 1-u \right)
\left\{ \tfrac{1}{k} \, d\tilde{\tau} - \beta \, d\varphi_{1} \right\}^{2} 
+ 4 \left( 1-\alpha^{2} \right) \left\{ \tfrac{1}{k} \, d\tilde{\tau} + \left( 1-\beta \right) d\varphi_{1} \right\}^{2} \\
\nonumber && \hspace{-0.25cm} \hspace{1.025cm} + \, 4 \hspace{0.025cm} \alpha^{2} u \left( 1-z \right)
\left\{\tfrac{1}{k} \, d\tilde{\tau} - d\chi - \beta \, d\varphi_{1} \right\}^{2} 
+ 4 \hspace{0.025cm} \alpha^{2} u \hspace{0.025cm} z
\left\{\tfrac{1}{k} \, d\tilde{\tau} - d\chi - d\varphi_{2} - \beta \, d\varphi_{1} \right\}^{2},
\end{eqnarray}
in terms of the new radial coordinates.  The 7-form field strength is given by
\begin{equation}
F^{(7)} = 96 \hspace{0.05cm} \tfrac{1}{k} \hspace{0.05cm} R^{6} \hspace{0.05cm} \alpha^{5} u \hspace{0.15cm} d\alpha \wedge du \wedge dz 
\wedge d\tilde{\tau} \wedge d\chi \wedge d\varphi_{1} \wedge d\varphi_{2} = dC^{(6)},
\end{equation}
associated with the following 6-form potential:
\begin{equation}
C^{(6)} 
= 16  \hspace{0.05cm} \tfrac{1}{k} \hspace{0.05cm} R^{6} \hspace{0.05cm} \alpha^{6} u \hspace{0.15cm} 
du \wedge dz \wedge d\tilde{\tau} \wedge d\chi \wedge d\varphi_{1} \wedge d\varphi_{2}.
\end{equation}
The constant form of integration has been chosen such that $C^{(6)}$ vanishes when $\alpha = 0$.

\subsection{M5-brane orbifold giant graviton solution} 

Let $\alpha$ be constant and choose the M5-brane worldvolume coordinates $\sigma^{a} = (t,u,z,\sigma,\chi,\varphi_{2})$.
The direction of motion is 
\begin{equation} \hspace{-0.5cm}  
\chi_{2} = \tfrac{\tilde{\tau}}{k} + (1-\beta) \, \varphi_{1} = \tfrac{1}{2} \hspace{0.05cm} \xi(t) \hspace{0.5cm}
\end{equation}
and we shall take the ans\"{a}tze (\ref{angular-ansatz}) and (\ref{definition-radial-coordinates}) for the angular and radial coordinates.

The M5-brane action takes the form
\begin{equation} \label{M5-action-generic}
S_{\text{M5}} = -\frac{1}{(2\pi)^{5}} \int{d^{6}\sigma} \hspace{0.1cm} \sqrt{|\det{ g_{ab}}| \hspace{0.05cm}} \hspace{0.1cm} \pm 
\frac{1}{(2\pi)^{5}}\int \hspace{0.1cm} d^{6}\sigma \hspace{0.2cm} \epsilon^{a_{1} \cdots \hspace{0.05cm} a_{6}} \hspace{0.1cm} C^{(6)}_{a_{1}\cdots \hspace{0.05cm} a_{6}}
\end{equation}
in the absence of any worldvolume gauge fields.  Substituting our ansatz into this action, we obtain
\begin{equation} \label{action - sphere giant}
S_{\text{M5}} = \int{dt} \hspace{0.25cm} L, \hspace{0.5cm} \text{with} \hspace{0.3cm}
L = - \frac{kN}{2}  \left\{ \alpha^{5} \sqrt{\alpha^{2} + \left( 1 - \alpha^{2} \right)\left(1 - \dot{\xi}^{2}\right)} - \alpha^{6} \hspace{0.075cm} \dot{\xi} \right\} \hspace{0.4cm}
\end{equation}
the Lagrangian.  Here $N \equiv \tfrac{2R^{6}}{k\pi^{2}}$ is the flux through the $S^{7}/\mathbb{Z}_{k}$ space, which is dual to the rank of the ABJM gauge group.

\begin{figure}[htb!] 
\begin{center}
\includegraphics[width = 9.5cm,height = 7.0cm]{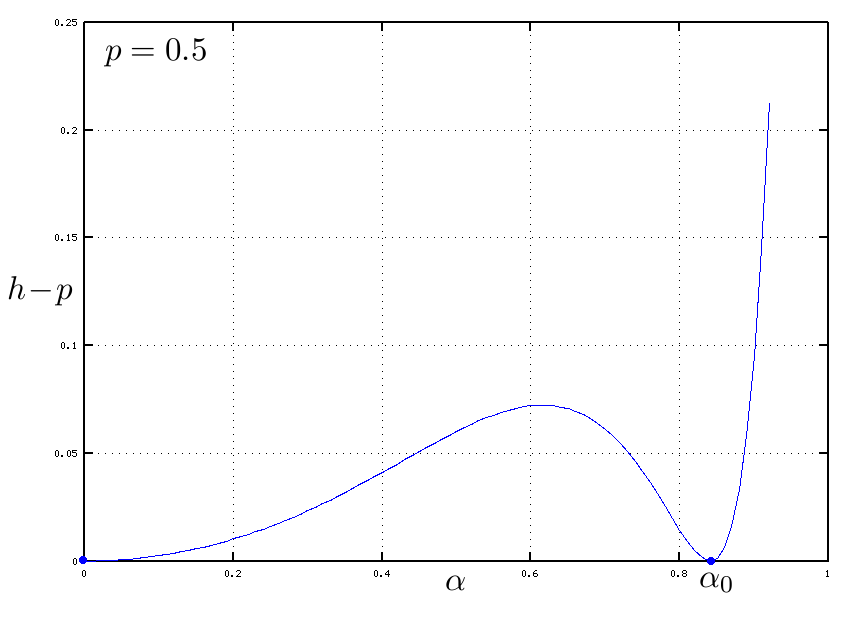}
\caption{The (shifted) energy of the M5-brane configuration in units of $\tfrac{kN}{2}$, given by $h-p$, plotted as a function of the (scaled) radius $\alpha \in [0,1]$ of the M5-brane at fixed $p = 0.5$.  The $\alpha = \alpha_{0} = p^{1/4}$ finite size, degenerate minimum is clearly evident.}
\label{energy plots}
\end{center}
\end{figure}

The momentum conjugate to $\xi$ is given by
\begin{equation}
P_{\xi} \equiv \frac{kN}{2} \hspace{0.1cm} p = \frac{kN}{2} \left\{ \frac{\alpha^{5} \left(1 - \alpha^{2} \right) \dot{\xi}}{\sqrt{\alpha^{2} + \left( 1 - \alpha^{2} \right)\left(1 - \dot{\xi}^{2}\right)}} + \alpha^{6} \right\},
\end{equation}
while the energy $H = P_{\xi} \hspace{0.1cm} \dot{\xi} - L$ can be written as
\begin{equation}
H \equiv \frac{kN}{2} \hspace{0.1cm} h =  \frac{kN}{2} \frac{\alpha^{5}}{\sqrt{\alpha^{2} + \left( 1 - \alpha^{2} \right)\left(1 - \dot{\xi}^{2}\right)}}.
\end{equation} 
Inverting the expression for the momentum as a function of the angular velocity gives
\begin{equation}
\left(1 - \alpha^{2}\right) \dot{\xi}^{2} = \frac{\left(p-\alpha^{6}\right)^{2}}{\left[ \left(p-\alpha^{6}\right)^{2} + \alpha^{10}\left(1 - \alpha^{2} \right) \right]}
\end{equation}
and hence the energy can be written as a function of the momentum and the size $\alpha$ as follows:
\begin{equation}
H = \frac{kN}{2} \frac{1}{\sqrt{1 - \alpha^{2}}} \sqrt{\left(p-\alpha^{6}\right)^{2} + \alpha^{10}\left(1 - \alpha^{2} \right)}.
\end{equation}
Plots of the energy at constant angular momentum are shown in figure \ref{energy plots}.  The finite size $\alpha = \alpha_{0} = p^{1/4}$ minimum (the giant graviton) occurs when $\dot{\xi} = 1$ and is energetically degenerate with the $\alpha = 0$ solution (the point graviton).  Notice that $H = P_{\xi}$ for the giant graviton solution, indicating a BPS configuration, as expected.  
\vspace{-0.2cm}

\section{$\tfrac{1}{2}$-BPS $\mathbb{CP}^{3}$ descendants} \label{section - half-BPS CP3 giants}
 
\subsection{$\mathbb{R} \times \mathbb{CP}^{3}$ background} 

We parameterize the $\mathbb{R}\times \mathbb{CP}^{3} \subset AdS_{4} \times \mathbb{CP}^{3}$ subspace, into which the D4-brane is embedded, in terms of the anti-de Sitter time $t$, the 
angular coordinates $(\chi,\varphi_{1},\varphi_{2})$ and the radial coordinates $(\alpha,u,z)$.  The metric is given by
\begin{equation}
ds^{2} = L^{2} \left\{ -dt^{2} + 4 \hspace{0.05cm} ds_{\mathbb{CP}^{3}}^{2} \right\} = L^{2} \left\{ -dt^{2} + ds_{\textrm{rad}}^{2} + ds_{\textrm{ang}}^{2} \right\}
\end{equation}
where we define
\begin{eqnarray}
&& \hspace{-0.5cm} ds_{\textrm{rad}}^{2} =  \frac{4 \hspace{0.075cm} d\alpha^{2}}{\left( 1-\alpha^{2} \right)} + \frac{\alpha^{2} \hspace{0.05cm} du^{2} }{u\left( 1-u \right)} + \frac{\alpha^{2} u  \hspace{0.1cm} dz^{2}}{z\left(1-z\right)} \\
\nonumber && \\
\nonumber && \hspace{-0.5cm} ds_{\textrm{ang}}^{2} = 4 \hspace{0.05cm} \alpha^{2}u\left(1-\alpha^{2}u\right)
\left[d\chi + \frac{\left( 1-\alpha^{2} \right)}{\left(1-\alpha^{2}u\right)} \hspace{0.075cm} d\varphi_{1} + z \hspace{0.075cm} d\varphi_{2}\right]^{2}  \\
&& \hspace{-0.5cm} \hspace{1.35cm} + \, \frac{4 \hspace{0.025cm} \alpha^{2}\left( 1-\alpha^{2} \right) \left( 1-u \right) }{\left(1-\alpha^{2}u \right)} \hspace{0.1cm} d\varphi_{1}^{2} 
+ 4\hspace{0.025cm} \alpha^{2}u \hspace{0.025cm} z \left(1-z\right) \hspace{0.05cm} d\varphi_{2}^{2}
\end{eqnarray}
and the dilaton $\Phi$ satisfies $e^{2\Phi} = \frac{4L^{2}}{k^{2}}$. 
The field strength forms in this subspace are
\begin{eqnarray}
&& \hspace{-1.5cm} F^{(2)} = k \hspace{0.15cm} d\mathcal{A} = dC^{(1)} \\
&& \hspace{-1.5cm} F^{(6)} = - 24 \hspace{0.05cm} k L^{4}  \hspace{0.05cm} \alpha^{5} u \hspace{0.15cm} d\alpha \wedge du \wedge dz \wedge d\chi \wedge d\varphi_{1} \wedge d\varphi_{2} = dC^{(5)} 
\end{eqnarray}
associated with the potential forms
\begin{eqnarray}
&& \hspace{-0.5cm} C^{(1)} 
= \hspace{0.05cm} k \left\{ (1-\beta)\, d\varphi_{1} -\alpha^{2} \left[ u \hspace{0.1cm} d\chi +  d\varphi_{1} + uz \hspace{0.1cm} d\varphi_{2} \right] \right\} 
=  k \hspace{0.05cm} \mathcal{A} \hspace{0.6cm} \\
&& \hspace{-0.5cm} C^{(5)} = - \hspace{0.05cm} 4 \hspace{0.05cm} k L^{4}  \hspace{0.05cm} \alpha^{6} u \hspace{0.15cm} du \wedge dz \wedge d\chi \wedge d\varphi_{1} \wedge d\varphi_{2}. \label{C5-potential}
\end{eqnarray}
Here we have taken into account the shift $\tau \rightarrow \tau - \beta \hspace{0.05cm} \varphi_{1}$ (or $\tilde{\tau} \rightarrow \tilde{\tau} - k\beta \hspace{0.05cm} \varphi_{1}$ after the orbifold identification) in the eleventh fibre direction, discussed in section \ref{section - half-BPS orbifold giants}.

\subsection{D4 and NS5-brane $\mathbb{CP}^{3}$ descendant solutions} 

Our $\tfrac{1}{2}$-BPS family of M5-brane orbifold giant gravitons is described by an ansatz  (\ref{angular-ansatz}) with the angular coordinates $\tilde{\tau}(t,\sigma)$ and $\varphi_{1}(t,\sigma)$ generally dependent on time $t$ and a spatial worldvolume coordinate $\sigma$.  Here $\sigma$ plays the role of an isometric direction in the dimensional reduction of the M5-brane in the orbifold $S^{7}/\mathbb{Z}_{k}$ to a D4 or NS5-brane in the complex projective space $\mathbb{CP}^{3}$.  The non-trivial dependence of $\varphi_1$ on this direction forces us to take a more general ansatz in which $\partial_\sigma X^\mu\neq 0$ for $\mu\in \{0,1,\dots , 9\}$ for this reduction. We still impose the condition that these derivatives are $\sigma$-independent.
For $\beta \in [0,1)$, the M5-brane is wrapped on the eleventh fibre direction $\tilde{\tau}$ 
and thus becomes a D4-brane after the compactification of $S^{7}/\mathbb{Z}_{k}$ to $\mathbb{CP}^{3}$.  The D4-brane is described by an action in which $\sigma$ has become a transverse isometric direction.  When compared with the standard action for a D4-brane, this new action contains extra couplings to a worldvolume scalar, $c^{(0)}$, which forms an invariant field strength, $\mathcal{F}^{(1)}$, with the RR 1-form potential. The D4-brane can therefore carry D0-brane charge.  
For $\beta = 1$, there is motion along the eleventh direction $\tilde{\tau}(t)$, but no wrapping on $\tilde{\tau}$.  After the compactification, the M5-brane becomes an NS5-brane, which again contains the $\sigma$ direction as an isometric direction, in this case lying on its worldvolume. The NS5-brane therefore spans an effectively 4+1 dimensional worldvolume. The action that describes this NS5-brane is identical to the new action of the D4-brane with an isometric direction. In particular, additional couplings to the worldvolume scalar, $c^{(0)}$, and its invariant field strength, $\mathcal{F}^{(1)}$, are again found, indicating D0-brane charge dissolved on the NS5-brane worldvolume.

The reduction of the M5-brane action (\ref{M5-action-generic}) in $AdS_{4} \times S^{7}/\mathbb{Z}_{k}$ to a D4 or NS5-brane action in $AdS_{4} \times \mathbb{CP}^{3}$ is described in appendix \ref{appendix - actions}. As already mentioned, this new action
may be seen (for $\beta \in [0,1)$) as the action of a D4-brane with an isometric transverse direction or (for $\beta =1$) as the action describing an NS5-brane with an isometric worldvolume direction after a dimensional reduction upon $l^\mu=\delta^\mu_\sigma$.
Substituting the M5-brane orbifold giant graviton ansatz (\ref{angular-ansatz}) into the DBI action (\ref{new-action-DBI}) gives
\begin{equation} 
S^{\text{DBI}}_{\text{D4/NS5}} = - \frac{1}{(2\pi)^{4}} \int{d^5\sigma} \hspace{0.15cm} \frac{k^2\,\alpha}{2L} \hspace{0.1cm} \sqrt{\left| \det{h}_{ab} \right|} \, ,  \hspace{0.15cm}
\end{equation}
where
\begin{equation}
h_{ab} \equiv {\cal G}_{ab} \, + \, \frac{e^{2\phi}\, l^2}{l^2+e^{2\phi} \, \mathcal{S}^2}\, 
(\mathcal{F}^{(1)})_a \, (\mathcal{F}^{(1)})_b 
\end{equation}
with $l^2=g_{\sigma\sigma}$ and
\begin{equation}
{\cal G}_{ab}=g_{ab}-\frac{g_{a\sigma}g_{b\sigma}}{g_{\sigma\sigma}}\, , \hspace{0.35cm}
\mathcal{S} = \partial_\sigma\, c^{(0)} + i_l \hspace{0.025cm} C^{(1)}\, , \hspace{0.35cm}
\mathcal{F}^{(1)} \equiv dc^{(0)} +\left[ (C^{(1)})_{a}  -\mathcal{S} \, \, \frac{g_{a\sigma}}{g_{\sigma\sigma}}\right] d\sigma^a, 
\end{equation}
as derived in appendix \ref{appendix - actions}.  Here $\mathcal{F}^{(1)}$ is the invariant field strength associated with D0-branes `ending' on the D4 or NS5-brane, with $c^{(0)}$ a worldvolume scalar whose origin is the eleventh direction. Note that $\dot{c}^{(0)}$ arises from the motion $\dot{\tilde{\tau}}$ along the fibre and $\p_{\sigma}c^{(0)}$ is related to the twist $\p_{\sigma} \tilde{\tau}$ on this eleventh direction. The CS action (\ref{new-action-CS}) becomes\footnote{We choose the positive sign here for a brane (rather than an anti-brane) configuration.}
\begin{eqnarray} 
S^{\text{CS}}_{\text{D4/NS5}}= \pm \hspace{0.05cm} \frac{1}{(2\pi)^{4}} \int{ d^{5}\sigma \hspace{0.2cm} 
\epsilon^{a_{1}\cdots \hspace{0.05cm} a_{5}} \left\{   k\left(1-\beta \right) \hspace{0.1cm} (C^{(5)})_{a_{1} \cdots \hspace{0.05cm} a_{5}}
 +  \partial_{[a_1}c^{(0)} \, (C^{(5)}) _{\varphi_{1} a_{2} \cdots \hspace{0.05cm} a_{5}]} \right\} }. \hspace{0.9cm}
\end{eqnarray}

The $\tfrac{1}{2}$-BPS M5-brane orbifold giant graviton ansatz (\ref{angular-ansatz}) then yields the following ansatz for the D4 or NS5-brane $\mathbb{CP}^{3}$ descendant:  The radial coordinate $\alpha$ is set to a constant.  Then $\sigma^{a} = (t,u,z,\chi,\varphi_{2})$ are chosen to be either the worldvolume coordinates of the D4-brane (for $\beta \in [0,1)$) or the reduced worldvolume coordinates of the NS5-brane (for $\beta=1$) after a dimensional reduction on the isometric worldvolume direction $l^{\mu} = \delta^{\mu}_{\sigma}$. The second term in the CS action arises from an interior product with this Killing vector.  The direction of motion in the complex projective space is $\varphi_{1}(t) = \tfrac{1}{2} \, \xi(t) $ and the worldvolume scalar takes the form $c^{(0)}(t) \equiv \tfrac{1}{2} \, k \hspace{0.025cm} \beta \, \xi(t)$. We can then compute
\begin{eqnarray}
&& \nonumber 
\hspace{-1.2cm} \hspace{0.47cm} h_{tt} = L^{2}  \left[-1 + \left(1-\alpha^{2} \right) \dot{\xi}^{2} \right]  \hspace{1.8cm}
h_{uu} = L^{2} \left\{ \frac{\alpha^{2}}{u\left( 1-u \right)} \right\} \\
&& \nonumber 
\hspace{-1.2cm} \hspace{0.38cm} h_{zz} = L^{2}  \left\{ \frac{\alpha^{2}u}{z\left( 1-z \right)} \right\} \hspace{2.95cm}
h_{\chi\chi} = L^{2} \left\{ 4 \, \alpha^{2} u \left( 1 - u \right) \right\} \\
&& 
\hspace{-1.2cm} h_{\varphi_{2}\varphi_{2}} = L^{2} \left\{ 4 \, \alpha^{2} uz \left( 1 - uz \right) \right\} \hspace{1.83cm}
h_{\chi\varphi_{2}} = L^{2} \left\{ 4 \, \alpha^{2} uz \left(1-u\right) \right\},
\end{eqnarray}
as well as
\begin{eqnarray}
\nonumber & \epsilon^{a_{1}\cdots \hspace{0.05cm} a_{5}} \left\{   k\left(1-\beta \right) \hspace{0.1cm} (C^{(5)})_{a_{1} \cdots \hspace{0.05cm} a_{5}}
 +  \partial_{[a_1} c^{(0)} \, (C^{(5)})_{\varphi_{1} a_{2} \cdots \hspace{0.05cm} a_{5}]} \right\} 
& = -\tfrac{1}{2} \hspace{0.1cm} k \, \dot{\xi} \hspace{0.1cm} (C^{(5)})_{u z \chi \varphi_{1} \varphi_{2}} \\
&& = 2 k L^{4} \alpha^{6} u \, \dot{\xi}. \hspace{2.5cm}
\end{eqnarray}

The D4 or NS5-brane action can now be simplified as follows:
\begin{equation} \label{action-D4-final}
S_{\text{D4/NS5}} = \int{dt} \hspace{0.25cm} L, \hspace{0.35cm} \text{with} \hspace{0.25cm}
L = - \frac{kN}{2}  \left\{ \alpha^{5} \sqrt{\alpha^{2} + \left( 1 - \alpha^{2} \right)\left(1 - \dot{\xi}^{2}\right)} - \alpha^{6} \hspace{0.075cm} \dot{\xi} \right\}. \hspace{0.2cm}
\end{equation}
This is identical in form to the action of the M5-brane orbifold giant graviton from which it descends.  Here $N \equiv \tfrac{kL^{4}}{2\pi^{2}}$ is the flux through the $\mathbb{CP}^{3}$ compact space.  The energy plot shown in figure \ref{energy plots}
is still applicable.  Now the momentum $P_{\xi} \equiv \tfrac{kN}{2} \hspace{0.05cm} p$ is fixed and the energy is expressed in units of $\tfrac{kN}{2}$. The $\alpha = \alpha_{0} = p^{1/4}$ minimum  occurs when $\dot{\xi} = 1$ and $\dot{c}^{(0)} = \frac{k \hspace{0.025cm} \beta}{2}$. Note that $ \tfrac{1}{2} \hspace{0.025cm} P_{\varphi_{1}} = (1-\beta) \, P_{\xi}$ and $\tfrac{1}{2} \hspace{0.05cm} k  \hspace{0.05cm} P_{c^{(0)}} = P_{\xi}$, so the parameter $\beta$ is related to the momentum in the complex projective space.  This momentum $P_{\varphi_{1}}$ vanishes precisely when $\beta=1$, being the case of the NS5-brane, which is supported only by D0-brane charge and not by angular momentum.
Since it is possible to view these D4 or NS5-brane $\mathbb{CP}^{3}$ descendants as M5-brane orbifold giant gravitons in the large $k$ limit, we
maintain that this remains a $\tfrac{1}{2}$-BPS configuration, as the above action would also seem to suggest.

\section{Discussion} \label{section - discussion}

This article is a comprehensive study of M5-brane giant gravitons and their various D4 and NS5-brane descendents. In addition to a survey of the literature of known giants, we have also shown the existence of several new classes of giant gravitons, both in the orbifold $AdS_{4}\times S^{7}/\mathbb{Z}_{k}$ as well as in its dimensional reduction to $AdS_{4}\times\mathbb{CP}^{3}$. A vital role in our constructions is played by Mikhailov's ingenious observation relating giant gravitons and holomorphic surfaces \cite{Mikhailov}. Here we have extended this holomorphic surface construction of giant gravitons to the ABJM correspondence by studying the effect of the orbifolding of $S^{7}$ by $\mathbb{Z}_{k}$ on the (moving) worldvolume $ \Sigma(t) = \mathcal{C}(t)\cap S^{7}$ of the M5-brane giant graviton. The direction of motion of this sphere giant graviton is the component of the preferred direction orthogonal to the worldvolume. While this preferred direction in $S^{7}$ induced by the complex structure of $\mathbb{C}^{4}$ is purely arbitrary, once made, it does break the $SO(8)$ isometry of the 7-sphere down to $SU(4)\cong SO(6)$. This choice is crucial, however, on the orbifold $S^{7}/\mathbb{Z}_{k} \subset \mathbb{C}^{4}/\mathbb{Z}_{k}$.  Here, the preferred direction of the giant graviton construction must now be chosen carefully in relation to the Hopf fibre of the orbifolding. Indeed, it is only for a particular choice of preferred direction that all the $\tfrac{1}{8}$-BPS sphere giant gravitons become $\tfrac{1}{6}$-BPS orbifold giant gravitons, with none of the configurations losing all its supersymmetry.  In this language then, the worldvolume of the orbifold giant graviton is $\Sigma(t)/\mathbb{Z}_{k} =  \mathcal{C}(t)/\mathbb{Z}_{k} \cap S^{7}/\mathbb{Z}_{k}$. To summarise our results, in addition to our generalization of Mikhailov's construction, we find and study in particular:
\begin{itemize}
\item{A new class of M5-brane orbifold giant gravitons embedded into $S^{7}/\mathbb{Z}_{k}$. These are generically $\tfrac{1}{6}$-BPS, but enjoy a supersymmetry enhancement to $\tfrac{1}{2}$-BPS and $\tfrac{1}{3}$-BPS configurations for certain simpler holomorphic surfaces.  These orbifold giant graviton solutions exhibit an $SU(4)$ symmetry. }
\item{A new class of D4 and NS5-branes, being $\mathbb{CP}^{3}$ descendants of the orbifold giant gravitons under the compactification of M-theory on $AdS_{4} \times S^{7}/\mathbb{Z}_k$ to type IIA superstring theory on $ AdS_{4} \times \mathbb{CP}^{3}$, which may be implemented as the $k\to\infty$ limit of the orbifold spacetime.  These configurations will have the same $\tfrac{1}{6}$-BPS (possibly enhanced to $\tfrac{1}{3}$-BPS or $\tfrac{1}{2}$-BPS) supersymmetry as their orbifold giant graviton ancestors. 
Included is a subclass of $\tfrac{1}{6}$-BPS D4-brane $\mathbb{CP}^{3}$ giant gravitons, supported only by their angular momentum in the complex projective space, with a supersymmetry enhancement to $\tfrac{1}{3}$-BPS configurations in special cases.  One such special case is the solution of \cite{GMP}.}
\item{A new, one-parameter family of $\tfrac{1}{2}$-BPS M5-brane orbifold giant gravitons embedded into $S^{7}/\mathbb{Z}_{k}$ as an explicit example of the above construction.  For this class of solutions the parameter $\beta \in [0,1]$ can be interpreted as a measure of how far away the direction of motion of the orbifold giant graviton is from the fibre direction (where $\beta=1$). Some properties of these solutions are:}
\begin{itemize}
\item{ For parameters $\beta \in [0,1)$, the orbifold giant graviton is moving and wrapped on the orbifold fibre, as well as moving in $\mathbb{CP}^{3}$. The embedding coordinates of the M5-brane are dependent on this eleventh fibre direction and its $\mathbb{CP}^{3}$ descendants are  D4-branes with both D0-brane charge, $P_{c^{(0)}} = 2 \hspace{0.025cm} k^{-1} P_{\xi}$, and angular momentum, $ P_{\varphi_{1}} = 2  \hspace{0.025cm} (1-\beta)  \hspace{0.025cm} P_{\xi}$, in the complex projective space.}
\item{ For the special case of $\beta=1$, with no wrapping along the eleventh direction, the $\mathbb{CP}^{3}$ descendant of the M5-brane orbifold giant graviton is an NS5-brane carrying D0-brane charge, 
$ P_{c^{(0)}} = 2  \hspace{0.025cm} k^{-1} P_{\xi}$, only.  This dielectric NS5-brane (NS5-D0-brane bound state) is the same as that found in  \cite{HLP} up to a change of worldvolume coordinates. }
\end{itemize}
  \item{ A new action describing D4 and NS5-branes with a special $U(1)$ isometric direction.
In the case of a D4-brane, this is transverse to the worldvolume, while for an NS5-brane it lies on the worldvolume. This action describes the $\tfrac{1}{2}$-BPS $\mathbb{CP}^{3}$ descendants in a unified way (both the D4 and NS5-branes), and contains an explicit coupling to a field strength, $\mathcal{F}^{(1)}$, constructed from the RR 1-form potential, which accounts for D0-brane charge in these brane configurations.}
\end{itemize}
We consider these configurations as an important step in the cataloguing of supersymmetric branes in M-theory and string theory, and hope that they will eventually find their place as indispensable entries in the AdS/CFT dictionary. Until then, there is much still left to do. Most pressing is, no doubt, to gain an understanding of their dual description in the gauge theory. We conclude this article with some thoughts on the dual ABJM operators as a topic for further study in the near future.

The operators dual to the most general orbifold giant gravitons and their $\mathbb{CP}^{3}$ descendants are not presently known. In the special case of D4-brane $\mathbb{CP}^{3}$ giant gravitons with no D0-brane charge, whose ans\"{a}tze lift directly from the complex projective space to the orbifold, there seems to be a clear link with operators in the ABJM model constructed from composite scalars of the form $A_{i}B_{j}$, as in \cite{Dey,dMMMP,Caputa:2012}, at least at zero coupling\footnote{As soon as one deviates from zero coupling $\lambda$, it becomes vital to consider the ordering of the ABJM scalar field components, $(\hat{A}_{i})^{a}_{~\beta}$ and $(\hat{B}_{j})^{~\alpha}_{b}$, in the dual operators and the problem gains an additional level of complexity.}. We anticipate that $\mathbb{CP}^{3}$ descendants with D0-charge will be dual to operators with additional monopole charge, such as those discussed in \cite{SJ,Berenstein-Park}.  In the case of the simplest $\tfrac{1}{2}$-BPS operators built from a single scalar field, $A_{i}$ or $B_{i}$, with invisible monopole operators attached to ensure gauge invariance, we suggest that linear combinations of the single trace operators of \cite{SJ} might be used to build Schur polynomials dual to the $\tfrac{1}{2}$-BPS orbifold giant gravitons and their $\mathbb{CP}^{3}$ descendants constructed in this work.  The requirement that the monopoles must be equivalent to large gauge transformations (and thus unobservable) implies that the ABJM operators should be built from the scalars $(A_{i})^{k}$ and $(B_{i})^{k}$ with monopoles attached. This is linked to the $\tfrac{1}{2}$-BPS orbifold giant gravitons being wrapped $k$ times on the eleventh fibre direction.  But that, as they say, is a whole other story$\ldots$

\section{Acknowledgements} \label{section - acknowledgements}
  
We would like to thank Jan Gutowski and Sanjaye Ramgoolam for useful discussions. Y.L. and J.M. are supported, in part, by the MPNS-COST Action MP1210 ``The String Theory Universe''.
J.M. also acknowledges the support of the National Research Foundation of South Africa under the Incentive Funding for Rated Researchers and HCDE programes. A.P. is grateful to the University of Oviedo for hospitality during an early stage of this project.

\appendix

\section{Backgrounds} \label{appendix - background} \label{appendix - backgrounds}

\subsection{$AdS_{4}\times S^{7}$ background} \label{appendix - backgrounds - sphere}

The $AdS_{4}\times S^{7}$ background is a maximally supersymmetric solution of the 11D SUGRA equations of motion.   In the Neveu-Schwarz sector, this solution has metric
\begin{equation} \label{metric-ads4xs7}
ds^{2} = R^{2} \left\{ ds_{AdS_{4}}^{2} + 4 \hspace{0.05cm} ds_{S^{7}}^{2} \right\},
\end{equation}
while the dilaton and B-field vanish.  The Ramond-Ramond field strength forms are
\begin{eqnarray}
&& F^{(4)} = - 3 R^{3} \hspace{0.075cm} \textrm{vol}(AdS_{4}) \\
&& F^{(7)} = \ast F^{(4)} =  3 (128) \hspace{0.05cm} R^{6} \hspace{0.075cm} \textrm{vol} (S^{7}).
\end{eqnarray}
Let us embed $S^{7} \subset \mathbb{C}^{4}$ into a complex manifold with coordinates:
 \begin{eqnarray}
\nonumber && z_{1} = r_{1} \hspace{0.05cm} e^{i\chi_{1}} 
= (2R) \hspace{0.05cm} \cos{\zeta} \sin{\tfrac{\theta_{1}}{2}} \hspace{0.1cm} e^{i\chi_{1}} \hspace{0.05cm}
= (2R) \hspace{0.05cm} \cos{\zeta} \sin{\tfrac{\theta_{1}}{2}} \hspace{0.1cm} e^{i\tau} \hspace{1.2cm} \\
\nonumber && z_{2} = r_{2} \hspace{0.05cm} e^{i\chi_{2}} 
= (2R) \hspace{0.05cm} \cos{\zeta} \cos{\tfrac{\theta_{1}}{2}} \hspace{0.1cm} e^{i\chi_{2}}
= (2R) \hspace{0.05cm} \cos{\zeta} \cos{\tfrac{\theta_{1}}{2}} \hspace{0.1cm} e^{i(\tau + \varphi_{1})} \hspace{1.2cm}  \\ 
\nonumber &&  z_{3} = r_{3} \hspace{0.05cm} e^{i\chi_{3}} 
= (2R) \hspace{0.05cm} \sin{\zeta} \sin{\tfrac{\theta_{2}}{2}} \hspace{0.1cm} e^{i\chi_{3}} \hspace{0.075cm}
= (2R) \hspace{0.05cm} \sin{\zeta} \sin{\tfrac{\theta_{2}}{2}} \hspace{0.1cm} e^{i(\tau - \chi)} \hspace{1.25cm} \\
&& z_{4} = r_{4} \hspace{0.05cm} e^{i\chi_{4}} 
= (2R) \hspace{0.05cm} \sin{\zeta} \cos{\tfrac{\theta_{2}}{2}} \hspace{0.1cm} e^{i\chi_{4}} \hspace{0.025cm}
= (2R) \hspace{0.05cm} \sin{\zeta} \cos{\tfrac{\theta_{2}}{2}} \hspace{0.1cm} e^{i(\tau - \chi - \varphi_{2})}, \hspace{1.2cm} 
\end{eqnarray}
in terms of $\chi_{a} \in [0,2\pi]$ or, alternatively, in terms of $\tau,\chi,\varphi_{i} \in [0,2\pi]$ with $\tau$ the overall phase. The $AdS_{4}$ and $S^{7}$ metrics are then given by
\begin{eqnarray}
&& \hspace{-0.5cm} ds_{AdS_{4}}^{2} = - \left( 1 + r^{2} \right) dt^{2} + \frac{dr^{2}}{\left( 1+r^{2} \right)} + r^{2} \left( d\theta^{2} + \sin^{2}{\theta} \hspace{0.075cm} d\varphi^{2} \right) \\
&& \hspace{-0.5cm} ds_{S^{7}}^{2} = \left(d\tau + \mathcal{A} \right)^{2} + ds_{\mathbb{CP}^{3}}^{2}
\end{eqnarray}
with the metric of $S^{7}$ written as a Hopf fibration $S^{7} \hookleftarrow \mathbb{CP}^{3}$ over a complex projective space.  Here 
\begin{eqnarray}
&& \hspace{-0.35cm} \mathcal{A}  =  
- \sin^{2}{\zeta}  \hspace{0.15cm} d\chi + \cos^{2}{\zeta}\cos^{2}{\tfrac{\theta_{1}}{2}} \hspace{0.15cm} d\varphi_{1} - \sin^{2}{\zeta} \cos^{2}{\tfrac{\theta_{2}}{2}} \hspace{0.15cm} d\varphi_{2}  \hspace{1.2cm}
\end{eqnarray}
and
\begin{eqnarray}
\nonumber && \hspace{-0.5cm} ds_{\mathbb{CP}^{3}}^{2} = d\zeta^{2} + \cos^{2}{\zeta} \sin^{2}{\zeta}
\left[d\chi + \cos^{2}{\tfrac{\theta_{1}}{2}} \hspace{0.075cm} d\varphi_{1} + \cos^{2}{\tfrac{\theta_{2}}{2}} \hspace{0.075cm} d\varphi_{2}\right]^{2} \\
&& \hspace{-0.5cm} \hspace{1.4cm} + \hspace{0.075cm} \tfrac{1}{4} \cos^{2}{\zeta} \left(d\theta_{1}^{2} + \sin^{2}{\theta_{1}} \hspace{0.05cm} d\varphi_{1}^{2} \right) 
+ \tfrac{1}{4} \sin^{2}{\zeta} \left(d\theta_{2}^{2} + \sin^{2}{\theta_{2}} \hspace{0.05cm} d\varphi_{2}^{2} \right)
\end{eqnarray}
is the Fubini-Study metric of the complex projective space $\mathbb{CP}^{3}$, parameterized by the radial coordinates $(\zeta,\theta_{i})$ and angular coordinates $(\chi,\varphi_{i})$.  It is evident that there are two squashed 
$S^{2}$'s  with coordinates $(\theta_{i},\varphi_{i})$ embedded in $\mathbb{CP}^{3}$.  The additional radial coordinate $\zeta$ controls the radii of these two $S^{2}$'s.  The additional angular coordinate $\chi$ describes a 
fibre direction.  The field strength forms are explicitly given by
\begin{eqnarray}
&& \hspace{-1.5cm} F^{(4)} = -3R^{3} r^{2} \sin{\theta} \hspace{0.15cm} dt \wedge dr \wedge d\theta \wedge d\varphi \\
&& \hspace{-1.5cm} F^{(7)} = 24 \hspace{0.05cm} R^{6} \sin^{3}{\zeta} \cos^{3}{\zeta} \sin{\theta_{1}} \sin{\theta_{2}} \hspace{0.15cm} 
d\zeta \wedge d\theta_{1} \wedge d\theta_{2} \wedge d\tau \wedge d\chi \wedge d\varphi_{1} \wedge d\varphi_{2}.
\end{eqnarray}

\subsection{$AdS_{4}\times S^{7}/\mathbb{Z}_{k}$ background} \label{appendix - backgrounds - orbifold}

To obtain the orbifold $AdS_{4}\times S^{7}/\mathbb{Z}_{k}$ solution of the 11D SUGRA equations of motion, we simply identify $\tau \sim \tau + \tfrac{2\pi}{k}$.  Defining $\tilde{\tau} = k \tau$ (where $\tilde{\tau}$ is an angular coordinate with the 
usual $2\pi$ periodicity), the background metric becomes
\begin{equation}
ds^{2} = R^{2} \left\{ ds_{AdS_{4}}^{2} + 4 \hspace{0.05cm} ds_{S^{7}/\mathbb{Z}_{k}}^{2} \right\},
\hspace{0.5cm} \text{with} \hspace{0.35cm}
ds_{S^{7}/\mathbb{Z}_{k}}^{2} = \frac{1}{k^{2}} \left(d\tilde{\tau} + k\mathcal{A} \right)^{2} + ds_{\mathbb{CP}^{3}}^{2}, \hspace{0.25cm}
\end{equation}
and the non-vanishing field strength forms are
\begin{eqnarray}
&& \hspace{-1.5cm} F^{(4)} =  - 3R^{3} \hspace{0.075cm} \textrm{vol}(AdS_{4}) 
 = -3R^{3} r^{2} \sin{\theta} \hspace{0.15cm} dt \wedge dr \wedge d\theta \wedge d\varphi \\
\nonumber && \hspace{-1.5cm} F^{(7)} 
= \ast F^{(4)} = 3 (128) \hspace{0.05cm} R^{6} \hspace{0.075cm} \textrm{vol}(S^{7}/\mathbb{Z}_{k}) \\
&& \hspace{-1.5cm} \hspace{0.84cm} 
= 24 \hspace{0.05cm} \tfrac{1}{k} \hspace{0.05cm} R^{6} \sin^{3}{\zeta} \cos^{3}{\zeta} \sin{\theta_{1}} \sin{\theta_{2}} \hspace{0.15cm} 
d\zeta \wedge d\theta_{1} \wedge d\theta_{2} \wedge d\tilde{\tau} \wedge d\chi \wedge d\varphi_{1} \wedge d\varphi_{2}.
\end{eqnarray}
This orbifold background solution retains 24 of the original 32 solutions of the Killing-Spinor equations, and is hence no longer maximally supersymmetric.

\subsection{$AdS_{4} \times \mathbb{CP}^{3}$ background} \label{appendix - backgrounds - CP3}

The $AdS_{4} \times \mathbb{CP}^{3}$ solution of the IIA 10D SUGRA equations of motion can be obtained from the $AdS_{4} \times S^{7}/\mathbb{Z}_{k}$ 11D SUGRA solution \cite{Pope-et-al:1984,Pope-et-al:1997} via a Kaluza-Klein reduction 
on an $S^{1}$ described by the Hopf fibre $\tilde{\tau}$:
\begin{eqnarray}
\nonumber & ds^{2} & = R^{2} \left( ds_{AdS_{4}} + 4 \hspace{0.05cm} ds_{\mathbb{CP}^{3}}^{2} \right) + \tfrac{4}{ k^{2} } \hspace{0.05cm} R^{2}  \left(d\tilde{\tau} + k \hspace{0.025cm} \mathcal{A} \right)^{2}  \\
&& = e^{-\frac{2}{3} \Phi} \hspace{0.1cm} L^{2} \left\{ ds_{AdS_{4}}^{2} + 4 \hspace{0.05cm} ds_{\mathbb{CP}^{3}}^{2} \right\} + e^{\frac{4}{3} \Phi} \left( d\tilde{\tau} + k \hspace{0.025cm} \mathcal{A} \right)^{2},
\end{eqnarray}
which implies: $ R^{3} = \tfrac{1}{2} k L^{2}$ and $e^{2\Phi} = \frac{4L^{2}}{k^{2}}$.  Moreover, $F_{4}$ has no legs in the $\tilde{\tau}$ direction, and hence $F_{4}$ and $B_{2} = 0$ remain unchanged by the reduction.

The AdS$_{4} \times \mathbb{CP}^{3}$ solution has the metric
\begin{equation} \label{metric-ads4xcp3}
ds^{2} = L^{2} \left\{ ds_{AdS_{4}}^{2} + 4 \hspace{0.05cm} ds_{\mathbb{CP}^{3}}^{2} \right\},
\end{equation}
with now a constant dilaton $\Phi$ in the Neveu-Schwarz sector.  The Ramond-Ramond field strength forms are given by
\begin{eqnarray}
\nonumber && \hspace{-1.5cm} F^{(2)} = k  \hspace{0.075cm} d\mathcal{A} = k \left\{ - \hspace{0.05cm} 2 \cos{\zeta} \sin{\zeta} \hspace{0.15cm} d\zeta \wedge 
\left[ d\chi + \cos^{2}{\tfrac{\theta_{1}}{2}} \hspace{0.1cm} d\varphi_{1} + \cos^{2}{\tfrac{\theta_{2}}{2}} \hspace{0.1cm} d\varphi_{2} \right]
\right. \\
&& \hspace{-1.5cm} \hspace{3.2cm} \left.
- \hspace{0.05cm} \tfrac{1}{2} \hspace{0.025cm} \cos^{2}{\zeta} \sin{\theta_{1}} \hspace{0.15cm} d\theta_{1} \wedge d\varphi_{1} + \tfrac{1}{2} \hspace{0.05cm} \sin^{2}{\zeta} \sin{\theta_{2} \hspace{0.15cm} d\theta_{2} \wedge d\varphi_{2}} \right\} \\
&& \hspace{-1.5cm} F^{(4)} = - \tfrac{3}{2} \hspace{0.05cm} k L^{2} \hspace{0.075cm} \textrm{vol}(AdS_{4}) = - \tfrac{3}{2} \hspace{0.05cm} k L^{2} \hspace{0.05cm} r^{2} \sin{\theta} \hspace{0.15cm} dt \wedge dr \wedge d\theta \wedge d\varphi  \\
\nonumber && \hspace{-1.5cm} F^{(6)} = \ast F^{(4)} = \tfrac{3}{2} (64) \hspace{0.05cm} k L^{4} \hspace{0.075cm} \textrm{vol}(\mathbb{CP}^{3}) \\
&& \hspace{-1.5cm} \hspace{0.84cm}= 6 \hspace{0.05cm} kL^{4}  \cos^{3}{\zeta} \sin^{3}{\zeta} \sin{\theta_{1}} \sin{\theta_{2}} \hspace{0.15cm} 
d\zeta \wedge d\theta_{1} \wedge d\theta_{2} \wedge d\chi \wedge d\varphi_{1} \wedge d\varphi_{2} \ \\
&& \hspace{-1.5cm} F^{(8)} = \ast F^{(2)}.
\end{eqnarray}
This $AdS_{4} \times \mathbb{CP}^{3}$ background can be thought of as the large $k$ limit of the orbifold $AdS_{4} \times S^{7}/\mathbb{Z}_{k}$ background and also preserves 24 of 32 possible supersymmetries.

\section{Details of the holomorphic surface construction} \label{appendix - review}

Here we review the construction \cite{Mikhailov} of a large class of $\tfrac{1}{8}$-BPS M5-brane giant gravitons embedded into $S^{7}\subset \mathbb{C}^{4}$ in the maximally supersymmetric 11D SUGRA geometry $AdS_{4}\times S^{7}$ from holomorphic surfaces in the complex manifold $\mathbb{C}^{4}$. 

\subsection{Holomorphic surfaces} \label{appendix - review - surfaces}

Let us define 
\begin{equation}
w_{a} \equiv \rho_{a} \hspace{0.05cm} e^{i \hspace{0.025cm} \psi_{a}} \equiv \left(2R\right) u_{a} \hspace{0.05cm} e^{i \hspace{0.025cm} \psi_{a}}
\end{equation} 
to be four coordinates for the complex manifold $\mathbb{C}^{4}$ with metric
\begin{equation} \label{metric-C4-w}
ds^{2} = \sum_{a=1}^{4} dw_{a} ~ d\bar{w}_{a} = \sum_{a=1}^{4} \left( d\rho_{a}^{2} + \rho_{a}^{2} ~ d\psi_{a}^{2} \right) 
= \sum_{a=1}^{4} \left(2R\right)^{2} \left( du_{a}^{2} + u_{a}^{2} ~ d\psi_{a}^{2} \right)
\end{equation} 
and complex structure fixed by this initial choice of complex coordinates $w_{a}$.  Here the submanifold $S^{7}$ is obtained by setting $(2R)^{2} = \sum_{a} \rho_{a}^{2}$ to a constant. The tangent space of the complex manifold $\mathrm{T} \mathbb{C}^{4}$ contains a single unit vector $e^{\perp} = \tfrac{1}{2} \hspace{0.05cm} \p_{R}$ orthogonal to the tangent space of the submanifold $\mathrm{T}S^{7}$.  The complex structure of $\mathbb{C}^{4}$ then induces a preferred direction ${\bf e}^{\parallel} = \mathrm{I} \hspace{0.05cm} {\bf e}^{\perp} = \tfrac{1}{2R} \hspace{0.05cm} \p_{\gamma}$ in $\mathrm{T}S^{7}$, with $\gamma$ the overall phase of the complex coordinates $w_{a} \sim e^{i\gamma}$.

We now consider a holomorphic surface $\mathcal{C}$ defined by $f(w_{1},w_{2},w_{3},w_{4})=0$ in $\mathbb{C}^{4}$ with $f$ some holomorphic function.  The intersection of this holomorphic surface $\mathcal{C}$ with the submanifold $S^{7}$ describes the (spatial) worldvolume $\Sigma = \mathcal{C} \, \cap \, S^{7}$ of an M5-brane giant graviton\footnote{A collection of M5-brane giant gravitons is found if $\Sigma$ is a collection of disjoint surfaces in $S^{7}$.} embedded into $S^{7}\subset \mathbb{C}^{4}$ at time $t=0$. The surface $\Sigma$ is then boosted into motion by applying a (reverse) boost along the preferred direction
$w_{a} \rightarrow w_{a} \hspace{0.05cm} e^{- \frac{i}{2} \hspace{0.025cm} \dot{\xi} \hspace{0.025cm} t}$ to the complex coordinates before substituting them into the holomorphic function:  
\begin{equation}
\mathcal{C}(t): \hspace{0.5cm} f(w_{1} \hspace{0.05cm} e^{-\frac{i}{2} \hspace{0.025cm} \dot{\xi} \hspace{0.025cm} t},w_{2} \hspace{0.05cm} e^{-\frac{i}{2} \hspace{0.025cm} \dot{\xi} \hspace{0.025cm} t} ,w_{3} \hspace{0.05cm} e^{-\frac{i}{2} \hspace{0.025cm} \dot{\xi} \hspace{0.025cm} t},w_{4} \hspace{0.05cm} e^{-\frac{i}{2} \hspace{0.025cm} \dot{\xi} \hspace{0.025cm} t})=0, \hspace{0.8cm} \text{with}  \hspace{0.35cm} 
\dot{\xi} = \pm 1.
\end{equation}
Here we are boosting along the overall phase $\gamma \rightarrow \gamma + \tfrac{1}{2} \hspace{0.05cm} \dot{\xi} \hspace{0.05cm} t$, so that the boost velocity is $v^{\parallel} = (2R)( \tfrac{1}{2} \hspace{0.05cm} \dot{\xi}) = R \hspace{0.05cm} \dot{\xi} $
on the great circle in $S^{7}$ parameterized by $\gamma$ and with radius $2R$.  This naturally leads to the motion of the surface $\Sigma$, not along the preferred direction ${\bf e}^{\parallel}$, but rather along the direction ${\bf e}^{\phi}$, which is the component of 
${\bf e}^{\parallel}$ orthogonal to the tangent space $\mathrm{T}\Sigma$:
\begin{equation}
{\bf e}^{\parallel} = -\cos{\mu} \hspace{0.15cm} {\bf e}^{\phi} \pm \hspace{0.05cm} \sin{\mu} \hspace{0.15cm} {\bf e}^{\psi}, 
\hspace{0.65cm} \text{with} \hspace{0.35cm} {\bf e}^{\psi} \in \mathrm{T} S^{7}.
\end{equation}
The velocity of the surface $\Sigma(t)$ along the direction of motion ${\bf e}^{\phi}$ is
$v^{\phi} = -R \hspace{0.05cm} \dot{\xi} \hspace{0.05cm} \cos{\mu}$, with $\mu \in [-\pi,\pi]$ a free parameter.

There is one other direction in $\mathrm{T}S^{7}$ (apart from the direction of motion ${\bf e}^{\phi}$)  orthogonal to $\mathrm{T}\Sigma$, since the surface $\Sigma$ is two real dimensions lower than the submanifold $S^{7}$.  This is defined to be a unit vector ${\bf e}^{n}$ which is also orthogonal to ${\bf e}^{\phi}$. In addition, we shall need the unit vector $\mathrm{I} \hspace{0.05cm} {\bf e}^{\phi}$, which is automatically orthogonal to both 
${\bf e}^{\phi}$ and $\mathrm{T}\Sigma$, and has components along both ${\bf e}^{n}$ (within $\mathrm{T}S^{7}$) and along ${\bf e}^{\perp}$ (orthogonal to $\mathrm{T}S^{7}$):
\begin{equation} \label{Iephi}
\mathrm{I} \hspace{0.05cm} {\bf e}^{\phi} = \cos{\mu} \hspace{0.15cm} {\bf e}^{\perp} + \hspace{0.05cm} \sin{\mu} \hspace{0.15cm} {\bf e}^{n}.
\end{equation}
Expressions for the unit vectors ${\bf e}^{\phi}$ and $\mathrm{I} \hspace{0.05cm} {\bf e}^{\phi}$ can be written down explicitly in terms of the derivatives $(\p_{w_{a}}f)$ of the holomorphic function, and the vectors 
$\p_{w_{a}}$ and $\p_{\bar{w}_{a}}$ in the tangent space $\mathrm{T}\mathbb{C}^{4}$ of the complex manifold:
\small
\begin{eqnarray}
& \nonumber {\bf e}^{\phi} & = \frac{1}{\left| \sum\limits_{c} \left(\p_{w_{c}}f\right) w_{c} \right|} 
\frac{1}{\sqrt{ \sum\limits_{d} \left| \p_{w_{d}}f \right|^{2} }} \\
&& \times \left\{ \left[ \sum_{a} \left(\p_{w_{a}}f\right) w_{a} \right] \sum_{b} \left(\overline{\p_{w_{b}}f}\right) \p_{w_{b}} 
-  \left[ \sum_{a} \left(\overline{\p_{w_{a}}f}\right) \bar{w}_{a} \right] 
\sum_{b} \left(\p_{w_{b}}f\right) \p_{\bar{w}_{b}} \right\}
\hspace{1.0cm}  \label{susy-exp01} 
\end{eqnarray}
\begin{eqnarray}
& \nonumber \mathrm{I} \hspace{0.05cm} {\bf e}^{\phi} & = \frac{1}{\left| \sum\limits_{c} \left(\p_{w_{c}}f\right) w_{c} \right|} 
\frac{i}{\sqrt{ \sum\limits_{d} \left| \p_{w_{d}}f \right|^{2} }} \\
&& \times \left\{ \left[ \sum_{a} \left(\p_{w_{a}}f\right) w_{a} \right] \sum_{b} \left(\overline{\p_{w_{b}}f}\right) \p_{w_{b}} 
+  \left[ \sum_{a} \left(\overline{\p_{w_{a}}f}\right) \bar{w}_{a} \right] 
\sum_{b} \left(\p_{w_{b}}f\right) \p_{\bar{w}_{b}} \right\}. \label{susy-exp02}
\hspace{1.0cm}  
\end{eqnarray}
\normalsize

The pullback of the $AdS_{4}\times S^{7}$ metric to the worldvolume of the giant graviton is then given by
\begin{equation}
ds^{2} = -\left\{R^{2} - (v^{\phi})^{2}\right\} \hspace{0.05cm} dt^{2} + \Sigma_{ij} \hspace{0.1cm} d\sigma^{i} d\sigma^{j} 
= -R^{2} \sin^{2}{\mu} \hspace{0.1cm} dt^{2} + \Sigma_{ij} \hspace{0.1cm} d\sigma^{i} d\sigma^{j},
\end{equation}
with $\Sigma_{ij}(t,\sigma^{k},\mu)$ the metric on the spatial worldvolume $\Sigma$(t) at time $t$. The volume element on the full worldvolume is 
\begin{equation}
\sqrt{|\det{g_{ab}}|} \hspace{0.25cm} dt \wedge d\sigma^{1} \wedge d\sigma^{2} \wedge d\sigma^{3} \wedge d\sigma^{4} \wedge d\sigma^{5} 
= R \left|\sin{\mu}\right| \hspace{0.1cm} dt \wedge \text{vol}\left(\Sigma\right). 
\end{equation}
It was argued in \cite{Mikhailov} that this M5-brane giant graviton satisfies a BPS bound $E=P_{\xi}$, indicating a supersymmetric configuration.

\subsection{Supersymmetry analysis}  \label{appendix - review - susy}

Let us now review the supersymmetry analysis of \cite{Mikhailov} for these M5-brane giant gravitons, which relies upon an embedding of the 11D curved spacetime $AdS_{4}\times S^{7}$ into the 13D flat spacetime $\mathbb{R}^{2+3} \times \mathbb{C}^{4}$.  We derive the KSE's and the kappa symmetry condition.

In this appendix, we will make use of the following convenient coordinate system for the $\mathbb{R}^{2+3} \times \mathbb{C}^{4}$ spacetime:
\begin{equation}
x^{\mu} = (t,r,\theta,\varphi,\hat{R}) \cup (\sigma^{1},\sigma^{2},\sigma^{3},\sigma^{4},\sigma^{5},x^{\phi},x^{n},2R),
\end{equation} 
where the metric takes the form
\begin{equation}
ds^{2} = -d\hat{R}^{2} + \hat{R}^{2} \, ds^{2}_{AdS_{4}} + d(2R)^{2} + (2R)^{2} \, ds^{2}_{S^{7}}  = g_{\mu\nu} \, dx^{\mu} \, dx^{\nu}.
\end{equation}
Here $g_{\mu\nu} = e^{\alpha}_{\mu} \, e^{\beta}_{\nu} \, \eta_{\alpha\beta}$ with signature $\eta = (-+++-)\cup (+ \, \cdots \, +)$ in terms of the vielbeins $e^{\alpha}_{\mu}$ in these coordinates.
The curved spacetime $\Gamma_{\mu} = e^{\alpha}_{\mu} \, \gamma_{\alpha}$ matrices can be written in terms of the flat space $\gamma_{\alpha}$ matrices.

The coordinates $\sigma^{i}$ run over the spatial worldvolume $\Sigma(t)$ of the M5-brane giant graviton, while $x^{\phi}$ parameterizes the direction of motion ${\bf e}^{\phi}$ and $x^{n}$ the orthogonal direction ${\bf e}^{n}$ also in $(T\mathcal{C})^{\perp}\cap \mathrm{T}S^{7}$. 
Note that $\p_{\hat{R}}$ is the unit vector in $\mathrm{T}\mathbb{R}^{2+3}$ orthogonal to $\mathrm{T}AdS_{4}$ and ${\bf e}^{\perp} = \p_{2R}$ is the unit vector in $\mathrm{T}\mathbb{C}^{4}$ orthogonal to $\mathrm{T}S^{7}$.  We make use of notation in which $\Gamma_{v} \equiv \Gamma({\bf e}^{v})$, where the coordinate $x^{v}$ is associated with some direction ${\bf e}^{v}$.  For the unit vectors ${\bf e}^{\phi}$, ${\bf e}^{n}$ and ${\bf e}^{\perp}$, we obtain
\begin{equation}
\gamma_{10} = \Gamma_{\phi} = \Gamma({\bf e}^{\phi}), \hspace{0.75cm} 
\gamma_{11} = \Gamma_{n} = \Gamma({\bf e}^{n})  \hspace{0.75cm} \text{and} \hspace{0.75cm}
\gamma_{12} = \Gamma_{2R} = \Gamma({\bf e}^{n}).
\end{equation}
We project out a 32-component Majorana spinor $\Psi$ from a 64-component complex spinor $\Psi_{+}$:
\begin{equation}
\Psi \equiv \tfrac{1}{2} \left( 1 - \gamma_{4} \gamma_{12} \right) \Psi_{+},
 \hspace{0.75cm} \text{so that} \hspace{0.5cm} \gamma_{4} \gamma_{12} \hspace{0.05cm} \Psi = - \Psi.
\end{equation}
Here $\hat{\gamma} \hspace{0.025cm} \gamma_{4} \hspace{0.025cm} \gamma \hspace{0.025cm} \gamma_{12} = 1$, since there is no chirality condition in 13D, with $\hat{\gamma} \equiv \gamma_{0} \gamma_{1} \gamma_{2} \gamma_{3}$ and $\gamma \equiv \gamma_{5}\gamma_{6}\gamma_{7}\gamma_{8}\gamma_{9}\gamma_{10}\gamma_{11}$ for convenience\footnote{Note again that our conventions for the flat spacetime metric imply $(\gamma_{0})^{2} = (\gamma_{4})^{2} = -1$ and $(\gamma_{1})^{2} = \cdots = (\gamma_{3})^{2} = (\gamma_{5})^{2} = \cdots = (\gamma_{11})^{2} = +1$.}.
We also impose the condition\footnote{The 64-component complex spinor $\Psi_{+}$ in 13D contains 128 degrees of freedom, while the 32-component Majorana spinor $\Psi$ in 11D onto which it is projected contains only 32 degrees of freedom.  We are therefore free to specify two consistent conditions on $\Psi_{+}$: (\ref{extra-spinor-condition01}), and later also (\ref{extra-spinor-condition02}).}
\begin{equation} \label{extra-spinor-condition01}
\gamma \hspace{0.025cm} \gamma_{12} \hspace{0.1cm} \Psi_{+} \equiv \Psi_{+}, 
\end{equation}
from which it follows that $\gamma_{4} \, \Psi_{+} = - \hspace{0.025cm} \hat{\gamma} \, \Psi_{+}$ and 
$\gamma_{12} \, \Psi_{+} = - \hspace{0.025cm} \gamma \, \Psi_{+}$.

The KSEs in flat $\mathbb{R}^{2+3}\times \mathbb{C}^{4}$ spacetime, $D_{\mu}\Psi_{+} = 0$, imply that
$\p_{\hat{R}} \Psi_{+} = \p_{2R} \Psi_{+} = 0$, together with the KSEs in the $AdS_{4}\times S^{7}$ spacetime:
\begin{equation}
\left( D_{\mu} + \frac{1}{2R} \hspace{0.15cm} \Gamma_{\mu} \,\hat{\gamma} \right) \Psi_{+} = 0 \hspace{0.75cm} \text{and} \hspace{0.75cm}
\left( D_{\mu} - \frac{1}{4R} \hspace{0.15cm} \Gamma_{\mu} \,\gamma \right) \Psi_{+} = 0,
\end{equation}
for $\mu \in \{0,\ldots,3\}$ and $\mu \in \{5,\ldots,11\}$, respectively. Here $D_{\mu} = \p_{\mu} + \frac{1}{4}(\Omega_{\mu})^{\alpha\beta}\gamma_{\alpha}\gamma_{\beta}$ are now the supercovariant derivatives of $AdS_{4}\times S^{7}$, with $\hat{R} = R$ taken to be constant.  The solutions of these  $\mathbb{R}^{2+3}\times \mathbb{C}^{4}$ background KSEs take the form
\begin{equation}
\Psi_{+}(x^{\mu}) = \mathcal{M}_{\mathbb{R}^{2+3}} \hspace{0.1cm}  \mathcal{M}_{\mathbb{C}^{4}} \hspace{0.2cm} \Psi_{+}^{0},
\end{equation}
where the $AdS_{4} \subset \mathbb{R}^{2+3}$ dependence of the solution is explicitly given by
\begin{equation} \label{KS-AdS-solution}
\mathcal{M}_{\mathbb{R}^{2+3}}(t,r,\theta,\varphi) 
= e^{- \frac{1}{2} \hspace{0.025cm} \rho \, \gamma_{1}\hat{\gamma}} \, e^{-\frac{1}{2} \hspace{0.025cm} t \, \gamma_{0}\hat{\gamma}} \,
e^{\frac{1}{2} \hspace{0.025cm} \theta \, \gamma_{12}} \, e^{\frac{1}{2} \hspace{0.025cm} \varphi \, \gamma_{23}} \hspace{0.5cm} \text{with} 
\hspace{0.25cm}r = \cosh{\rho},
\end{equation}
and the $\mathbb{C}_{4}$ dependence in $\mathcal{M}_{\mathbb{C}^{4}}$ is discussed in section \ref{section - sphere giants}.

The kappa symmetry condition for an M5-brane embedded into $S^{7}$ takes the form
\begin{equation}
\Gamma \, \Psi = \Psi \hspace{0.75cm} \text{with} \hspace{0.5cm} 
\Gamma \equiv -\frac{1}{6!} \frac{\epsilon^{i_{0}\cdots \hspace{0.05cm} i_{5}}}{\sqrt{|\det{g_{ab}}|}} \, 
\left( \p_{i_{0}} X^{\mu_{0}} \right) \cdots \left( \p_{i_{5}} X^{\mu_{5}} \right) \, \Gamma_{\mu_{0} \cdots \hspace{0.05cm} \mu_{5}}
\end{equation}
pulled back to the worldvolume $\mathbb{R} \times \Sigma(t)$. The sphere giant gravitons constructed from holomorphic surfaces  have $\sqrt{|\det{g_{ab}}|} = R \, |\sin{\mu}| \, \sqrt{\det \Sigma_{ab}}$ with worldvolume coordinates $(t,\sigma^{i})$. Here also 
$\dot{x}^{\phi} = v^{\phi} = - R \, \dot{\xi} \, \cos{\mu}$. Hence we can compute
\begin{equation}
\hspace{-1.0cm}
\Gamma = \frac{-1}{|\sin{\mu}|} \hspace{0.1cm}( \gamma_{0} - \dot{\xi} \, \cos{\mu} \hspace{0.1cm} \gamma_{10} ) \hspace{0.15cm}
\gamma_{10} \hspace{0.05cm} \gamma_{11} \hspace{0.05cm} \gamma_{12} \left(\gamma \hspace{0.025cm} \gamma_{12} \right),
\end{equation}
and the kappa symmetry condition can be rewritten as
\begin{equation}
\left(- \hspace{0.025cm} \dot{\xi} \, \gamma_{0} \gamma_{4} \right) \gamma_{10} \left\{ \dot{\xi} \, |\sin{\mu}| \, \gamma_{11} \left(\gamma \gamma_{12}\right) +  \cos{\mu} \, \gamma_{12} \right\} \Psi = \Psi.
\end{equation}
Notice that the operator on the left-hand side of the above expression commutes with the projection operator 
$\tfrac{1}{2} \left( 1 - \gamma_{4} \gamma_{12} \right)$ and hence this condition is satisfied for $\Psi$ if it is satisfied for $\Psi_{+}.$  Using the additional properties of $\Psi_{+}$, we shall thus insist that
\begin{equation}
\left(\dot{\xi} \, \gamma_{0} \hat{\gamma} \right) \gamma_{10} \left\{ \dot{\xi} \, |\sin{\mu}| \, \gamma_{11} + \cos{\mu} \, \gamma_{12} \right\} \Psi_{+} = \Psi_{+},
\end{equation}
with this spinor $\Psi_{+}$  pulled-back to the holomorphic surface $\mathbb{R}\times\mathcal{C}(t)$ embedded and moving in $\mathbb{R}^{2+3} \times \mathbb{C}^{4}$. 
Let us now choose $\mu \in [-\pi,\pi]$ such that $\dot{\xi} \, |\sin{\mu}| = \sin{\mu}$. We shall also impose the extra condition 
\begin{equation} \label{extra-spinor-condition02}
\gamma_{0} \hat{\gamma} \hspace{0.1cm} \Psi_{+} \equiv \, - \hspace{0.025cm} i \hspace{0.05cm} \dot{\xi} \hspace{0.1cm} \Psi_{+}.
\end{equation}
Here we have simply chosen the orientation of the M5-brane configuration and the eigenvalue of $\gamma_{0} \hat{\gamma}$ to be related to the direction of the motion $\dot{\xi} = \pm 1$ along ${\bf e}^{\phi}$.  Note that any Dirac bilinear in the $\mathbb{C}^{4}$ space (for example, $\gamma_{5}\gamma_{9}, \gamma_{6}\gamma_{10}$ and $\gamma_{7}\gamma_{11}$), as well as $\gamma_{1}\hat{\gamma}$ and $\gamma_{2}\gamma_{3}$, with eigenvalues which may be used to label the 32 degrees of freedom in the spinor $\Psi_{+}$, commute with $\gamma_{0}\hat{\gamma}$.  It is therefore sensible to impose this constraint. The kappa symmetry condition then reduces to
\begin{equation}
\gamma_{10} \left( \sin{\mu} \hspace{0.1cm} \gamma_{11} + \cos{\mu} \hspace{0.1cm} \gamma_{12} \right) \Psi_{+} = i \, \Psi_{+}
\end{equation}
and (\ref{Iephi}) allows us to rewrite it as
\begin{equation}
\Gamma({\bf e}^{\phi}) \hspace{0.1cm} \Gamma(\textrm{I} \, {\bf e}^{\phi}) \hspace{0.1cm} \Psi_{+} = i \, \Psi_{+}.
\end{equation}
From our expressions (\ref{susy-exp01}) and (\ref{susy-exp02}), we obtain
\begin{eqnarray}
\Gamma({\bf e}^{\phi}) \hspace{0.1cm} \Gamma(\mathrm{I} \hspace{0.05cm} {\bf e}^{\phi})  = i \hspace{0.05cm} + \hspace{0.05cm} 
\frac{2 \hspace{0.05cm} i}{ \sum\limits_{c} \left| \p_{w_{c}}f \right|^{2}} 
\sum_{a,b} \left(\overline{\p_{w_{a}}f}\right) \left(\p_{w_{b}}f \right) \hspace{0.1cm} \Gamma_{w_{a}}\hspace{0.05cm} \Gamma_{\bar{w}_{b}}. \hspace{0.9cm}
\end{eqnarray}
This kappa symmetry condition will be satisfied if we take
\begin{equation} \label{kappa-gamma-wa}
\Gamma_{\bar{w}_{a}} \hspace{0.05cm} \Psi_{+}  = 0
\end{equation}
for all the complex coordinates $w_{a}$ such that $(\p_{w_{a}} f) \neq 0$.  It is already apparent that we should expect supersymmetry enhancement for those giant gravitons associated with holomorphic functions which are independent of several complex coordinates.

\section{The D4 and NS5-brane actions} \label{appendix - actions}

Let us consider an M5-brane embedded into $AdS_{4} \times S^{7}/\mathbb{Z}_{k}$ with worldvolume coordinates $\sigma^{A}=(t,u,z,\chi,\varphi_{2},\sigma)\equiv(\sigma^{a},\sigma)$. The embedding coordinates $X^{M}$ are labeled by 
$M \in \{0,1,\ldots,10\}$.  
Here we take the eleventh direction $X^{10} = \tilde{\tau}(t,\sigma)$ to be dependent on the worldvolume time $t$ and on the spatial worldvolume coordinate $\sigma$, with the $\dot{\tilde{\tau}}$ and $\p_{\sigma}\tilde{\tau}$ derivatives $\sigma$-independent.
Under the compactification to $AdS_{4} \times \mathbb{CP}^{3}$, this M5-brane becomes either a D4-brane with worldvolume coordinates $\sigma^{a}$ (if $\tilde{\tau}$ depends on $\sigma$ so that the M5-brane is twisted on the eleventh direction) or an NS5-brane with worldvolume coordinates $\sigma^{A}$ (in the special case in which $\tilde{\tau}$ is independent of $\sigma$) with embedding coordinates $X^{\mu}$ labeled by $\mu \in \{ 0,1,\ldots,9 \}$. Due to the fact that, in our ansatz (\ref{angular-ansatz}), the $\mathbb{CP}^{3}$ angular coordinate $\varphi_1$ depends on $\sigma$, we are forced to make a more general reduction from eleven dimensions in which we allow $\partial_\sigma X^\mu\neq 0$. Still, we impose the condition that these derivatives are $\sigma$ independent. Then $\sigma$ (or, equivalently, $\varphi_1$) appears in the reduced action as a special isometric direction which is transverse to the D4-brane and lies in the worldvolume of the NS5-brane. In the D4-brane case, the resulting action thus differs from the standard DBI plus CS action in that it contains this special isometric direction in the transverse space. This allows, in fact, for the inclusion of an explicit coupling to the RR 1-form potential, which indicates D0-brane charge in the configuration. For the NS5-brane, in turn, we recover the action constructed in  \cite{HLP}  describing an NS5-brane with a $U(1)$ isometric worldvolume direction. When compared with the action of the unwrapped NS5-brane in type IIA superstring theory, constructed in \cite{BLO,EJL}, we see that the self-dual 2-form field of the latter is replaced by a vector, associated with D2-branes wrapping the isometric direction. This allowed for a closed form for the reduced action to be given.

With this reduction ansatz 
the M5-brane DBI action becomes
\begin{eqnarray} 
\label{new-action-DBI}
S^{\mathrm{DBI}}_{\mathrm{D4/NS5}}=-\frac{1}{(2\pi)^4} \int d^5\sigma  \hspace{0.15cm} e^{-2\phi}\sqrt{l^2+e^{2\phi}\, \mathcal{S}^2} \, \, 
\sqrt{ \left|\det{\left(P[{\cal G}]+\frac{e^{2\phi}\,l^2}{l^2+e^{2\phi}\, \mathcal{S}^2} \, ({\cal F}^{(1)})^2\right)}\right|}\, , \hspace{0.8cm}
\end{eqnarray}
for a vanishing worldvolume 3-form field strength. 
Here $l^\mu= \delta^{\mu}_{\sigma}$ is an Abelian Killing vector that points along the $\sigma$ direction and ${\cal G}$ is the reduced metric with components ${\cal G}_{\mu\nu}=g_{\mu\nu}-l^{-2}l_\mu l_\nu$, where $l_\mu=g_{\mu\nu}\,l^\nu$.  The scalar  $\mathcal{S}$ has been defined such that
$\mathcal{S} =\partial_\sigma\, c^{(0)}+i_l \hspace{0.025cm} C^{(1)}$, with $i_l \hspace{0.025cm} C^{(p)}$ the interior product of the $C^{(p)}$ potential with the Killing vector.
The worldvolume scalar field $c^{(0)}$ has its origin in the eleventh direction $\tilde{\tau}$ and forms an invariant field strength 
\begin{equation}
\mathcal{F}^{(1)}=dc^{(0)}+P[C^{(1)}]-\mathcal{S} \,\, P\,\left[\frac{l_1}{l^2}\right],\,
\end{equation}
when combined with the RR 1-form potential $C^{(1)}$ and the 1-form $l_1$ with $l_\mu$ components. It is therefore 
associated with D0-branes `ending' on the D4 or NS5-brane. Written in this form, the action is manifestly isometric under translations along the Killing direction. Note that the 4 transverse scalars, together with the worldvolume scalar $c^{(0)}$ and the vector field that arises upon reduction of the worldvolume 2-form field of the M5-brane (which we have omitted in this calculation), give the right counting of bosonic degrees of freedom, 8, on the 5-dimensional worldvolume.
In the case of the NS5-brane, since there is no wrapping on $\tilde{\tau}$, $\partial_\sigma\, c^{(0)}$ vanishes. In this case the DBI action (\ref{new-action-DBI}) coincides with the expression given in \cite{HLP} for vanishing 2-form field strength.
The reader is referred to \cite{HLP} for more details.

The M5-brane CS action reduces to
\begin{equation} \label{new-action-CS}
S^{\text{CS}}_{\text{D4/NS5}}= \pm \hspace{0.05cm} \frac{1}{(2\pi)^{4}} \int{ 
 \left\{  - P\, [i_l B^{(6)}]+ \p_{\sigma}\,  c^{(0)}\wedge P\, [ C^{(5)}] - dc^{(0)} \wedge  P\, [i_l C^{(5)} ]  \right\} },
\end{equation}
where we integrate over $\sigma^{a} = (t,u,z,\chi,\varphi_{2})$ and  $B^{(6)}$ denotes the Hodge dual of the NS-NS 2-form field. Note that, in the actions (\ref{new-action-DBI}) and (\ref{new-action-CS}), the pullback $P$ is taken with respect to the $\sigma^{a}$ worldvolume directions only. For example:
\begin{equation}
\hspace{-0.75cm}
 P\left[C^{(1)}\right] = C^{(1)}_{\mu} \left(\p_{a} X^{\mu}\right) \, d\sigma^{a}, \hspace{0.5cm}
P\left[\frac{l^{\mu}}{l^{2}}\right] = - \, \frac{g_{\sigma\mu}}{g_{\sigma\sigma}} \, \left( \p_{a} X^{\mu} \right) \, d\sigma^{a}, \hspace{0.5cm} \ldots\\
\end{equation}
We stress that this action is applicable to both D4 and NS5-branes:  For the D4-brane, we interpret $\delta^{\mu}_{\sigma}$ as the isometric transverse direction upon which the M5-brane ancestor was wrapped. For the NS5-brane, we interpret $\delta^{\mu}_{\sigma}$ as an isometric worldvolume direction of the M5-brane ancestor and of the NS5-brane descendant.


\end{document}